\documentclass[sigconf,screen]{acmart}	% camera-ready version

\setcopyright{acmlicensed}
\acmPrice{}
\acmDOI{10.1145/3503222.3507754}
\acmYear{2022}
\copyrightyear{2022}
\acmSubmissionID{asplos22main-p740-p}
\acmISBN{978-1-4503-9205-1/22/02}
\acmConference[ASPLOS '22]{Proceedings of the 27th ACM International Conference on Architectural Support for Programming Languages and Operating Systems}{February 28 -- March 4, 2022}{Lausanne, Switzerland}
\acmBooktitle{Proceedings of the 27th ACM International Conference on Architectural Support for Programming Languages and Operating Systems (ASPLOS '22), February 28 -- March 4, 2022, Lausanne, Switzerland}

\usepackage{balance}

\usepackage{microtype}

\usepackage{wrapfig}
\usepackage{comment}
\usepackage{colortbl}

\usepackage{color,soul}
\usepackage{graphics}
\usepackage{hyperref}
\usepackage{lipsum}
\usepackage{tikz}
\usepackage{enumitem}	% for fine control of list env
\usepackage{listings} % for code listing
\usepackage{booktabs}	% xzl: for fancy tables
\usepackage{lipsum}

\usepackage{capt-of}	% xzl: used to override the ``figure'' and ``table'' in caption.

\usepackage{todonotes}  %xzl: for "missing figures"
\usepackage[caption=false]{subfig}

\definecolor{labelkey}{rgb}{0,1,0}

\usepackage{mathrsfs}	% xzl: for Ralph Smith’s Formal Script Font (rsfs)
\usepackage{amsfonts}

\usepackage[export]{adjustbox} % xzl: flush figure to left/right

\usepackage{algorithm,algorithmic,amsmath} % jin: for pseudo code
\usepackage{boldline}					% jin: use boldline in the table

\usepackage{multirow}

\renewcommand{\paragraph}[1]{\vskip 3pt\noindent\textbf{#1 }}	 % used to be 6pt

\newenvironment{smitemize}%
  {\begin{list}{$\bullet$}%
     {\setlength{\parsep}{0pt}%
      \setlength{\topsep}{0pt}%
      \setlength{\itemsep}{2pt}}}%
  {\end{list}}
\newcommand\xzlNote[1]{\sethlcolor{yellow} \hl{#1}} % highlighted notes of other colors.
\newcommand\Noted[1]{} % remove highlights.

\definecolor{darkblue}{rgb}{0.0, 0.0, 0.55}
\definecolor{mygreen}{HTML}{ADFF2F}
\definecolor{mylightgray}{gray}{0.8}
\newcommand\jin[2][]{
	\fcolorbox{blue}{white}{\bf\em\color{blue}jin:}
	{\small\em\color{blue}{\fontfamily{qhv}\selectfont \underline{#1} #2}}
}

\newcommand\myfootnote[1]{\textcolor{red}{\footnote{#1}}}

\newenvironment{myitemize}%
  {\begin{itemize}
	[leftmargin=0cm,
		itemindent=.3cm,
		labelwidth=\itemindent,
		labelsep=0pt,
		parsep=1pt,
		topsep=1pt,
		itemsep=1pt,
		align=left]
  }%
  {\end{itemize}}    

\newenvironment{myenumerate}%
  {\begin{enumerate}
	[leftmargin=.cm,itemindent=.5cm,labelwidth=\itemindent,
		labelsep=0pt,
		parsep=1pt,
		topsep=1pt,
		itemsep=3pt,
		align=left]
  }%
  {\end{enumerate}}    

\newcommand\sect[1]{Section~\ref{sec:#1}}	% NB: does not play well with \note{}        

\newcommand{\code}[1]{\texttt{\small{#1}}}	
\newcommand{\sysfull}{GPUReplay}
\newcommand{\sys}{GR}
\newcommand{\rec}{TS-Recorder}
\newcommand{\mm}{memory manager}
\newcommand{\rn}{runner}
\newcommand{\disp}{dispatcher}
\newcommand{\rep}{TS-Replayer}

\makeatletter
\def\@copyrightspace{\relax}
\makeatother

\begin{document}

% ****************** TITLE *******************************************

%\title{TZGPU: Trustworthy Mobile GPU Execution with TrustZone}
%\title{TZGPU: A minimalist software stack for mobile GPU}
%\title{\sys{}: A 50-KB GPU Stack for Client Computation}
\title{\sysfull{}: A 50-KB GPU Stack for Client ML}
%\title{\sys{}: A minimalist GPU Stack for Client Computation}
 
% ****************** AUTHORS **************************************

\author{Heejin Park}
\affiliation{%
	\institution{Purdue University}
	\city{West Lafayette}
	\state{Indiana}
	\country{USA}
}
\email{bakhi@purdue.edu}

\author{Felix Xiaozhu Lin}
\affiliation{%
	\institution{University of Virginia}
	\city{Charlottesville}
	\state{Virginia}
	\country{USA}
}
\email{felixlin@virginia.edu}




% ----------------------------------------------
\date{}
% !TeX root = main.tex

\begin{abstract}

\sysfull{} (\sys{}) is a novel way for deploying GPU-accelerated computation on mobile and embedded devices. 
%Modern GPU stacks were architected for graphics workloads in the first place their complexity becomes a burden to compute in ML workloads. 
% It addresses the complexity of a modern GPU stack which is burdening compute workloads. 
It addresses high complexity of a modern GPU stack for deployment ease and security.
%A modern GPU stack was architected for dynamic graphics workloads in the first place. 
% It becomes a burden for ML which runs prescribed, coarse-grained GPU compute. 
%\st{Without an overhaul of the stack, \sys{} provides a static, fast path for an app to push its computation to GPU.}
The idea is to record GPU executions on the full GPU stack ahead of time and replay the executions on new input at run time. 
% without the stack at run time.
%\sys{} addresses key challenges towards making replaying pre-recorded GPU computations feasible, robust, and practical to use. 
We address key challenges towards making \sys{} feasible, sound, and practical to use. 
%capturing key CPU/GPU interactions and GPU states, working around proprietary GPU internals, and ensuring replay correctness. 
The resultant replayer is a drop-in replacement of the original GPU stack. 
It is tiny (50 KB of executable), 
% robust (playing tens of thousands of register accesses without divergence), 
robust (replaying long executions without divergence), 
portable (running in a commodity OS, in TEE, and baremetal), 
and quick to launch (speeding up startup by up to two orders of magnitude).  
% We have implemented \sys{} for multiple integrated GPUs and have produced  recordings of neural networks from a variety of ML frameworks and programming APIs. 
%We have implemented \sys{} and tested it with a variety of ML frameworks, GPU APIs, and integrated GPU hardware.
%The prototype of \sys{} successfully records and replays 1 NN training and 33 NN inferences.
We show that \sysfull{} works with a variety of integrated GPU hardware, 
GPU APIs, ML frameworks, and 33 neural network (NN) implementations for inference or training. 
The code is available at \code{https://github.com/bakhi/GPUReplay}.
%\xzlNote{XXXX}. 

% \Note{record once, replay anywhere} --- probably bad to claim

%\input{todo}

\end{abstract}
\maketitle
%\pagestyle{plain}

% !TeX root = main.tex

\section{Introduction}
\label{sec:intro}

%%%%%%%%%%%%%%%%%%%%%%%%%%%%%%%%%%%%%%%%%%%%%%%%%%%%%%%%%%%%%%%%%%%%%%%%%%%%%%%%%%%%%%%%%%%%%%%%%%%%%%%%
%%%%%%%%%% GPU is used as accelerator for security-sneitive workload in on-deivce computation %%%%%%%%%%
%%%%%%%%%%%%%%%%%%%%%%%%%%%%%%%%%%%%%%%%%%%%%%%%%%%%%%%%%%%%%%%%%%%%%%%%%%%%%%%%%%%%%%%%%%%%%%%%%%%%%%%%

\begin{figure}[t!]
\centering
\includegraphics[width=0.46\textwidth{}]{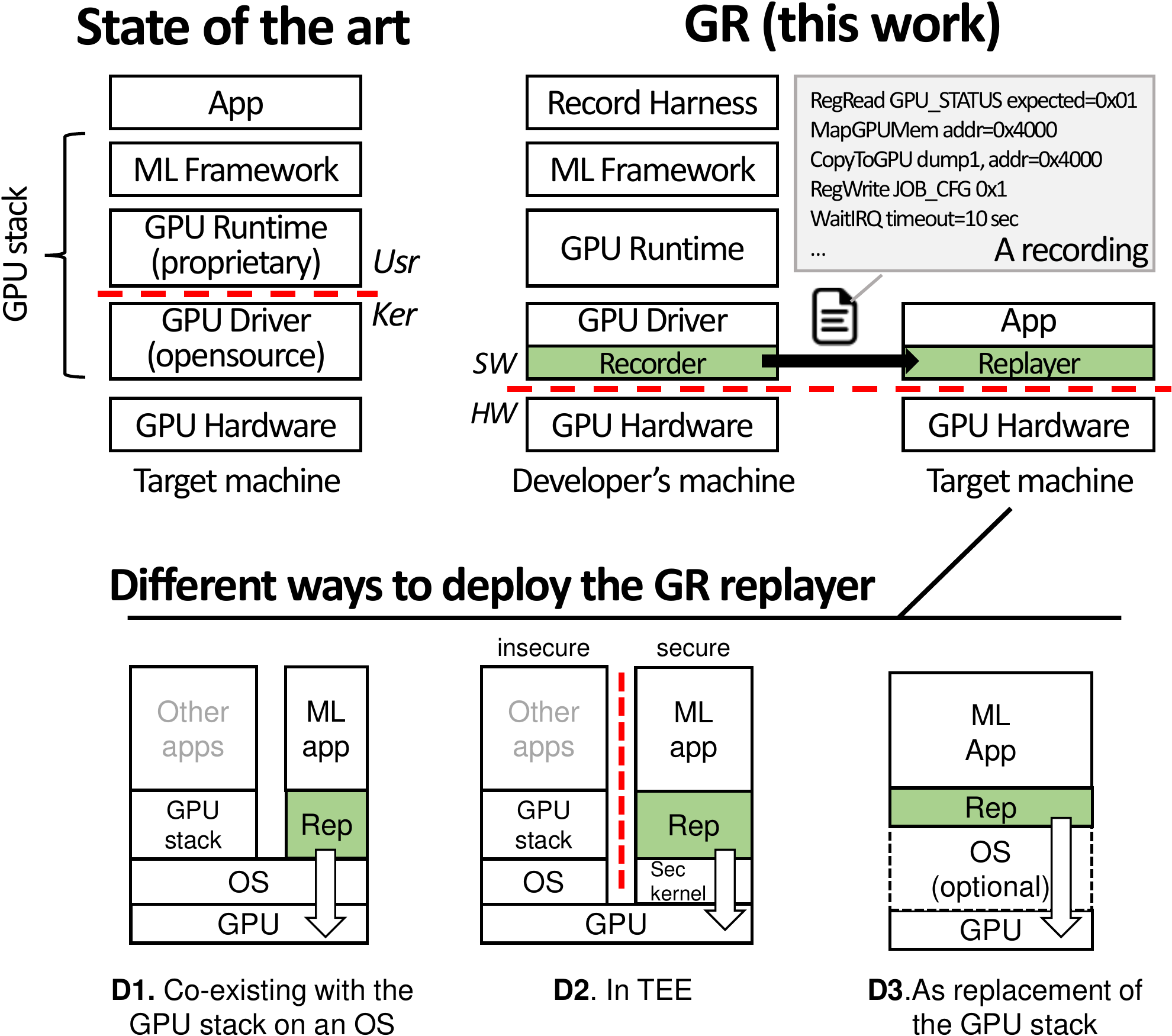} 
%\missingfigure[figwidth=6cm]{Juno platform power instrumentation}
\vspace{-3pt}		% use as needed
\caption{The overview of \sys{}}
\label{fig:overview}
\vspace{-10pt}		% use as needed
\end{figure}

\paragraph{GPU stacks}
%Client devices, e.g. 
Smartphones or IoT devices
commonly use GPUs to accelerate machine learning (ML). 
As shown in Figure~\ref{fig:overview},
a modern GPU software stack spans 
ML frameworks (e.g. Tensorflow~\cite{tensorFlow} and ncnn~\cite{ncnn}), a GPU runtime (e.g. OpenCL or Vulkan runtimes) that translates APIs to GPU commands and code, and a GPU driver that tunnels the resultant code and data to GPU. 
% the device driver set up data and control paths between GPU runtimes and GPU hardware;
% consisting millions lines of code resulting in of hundreds of MBs binary executables. \Note{numbers}
A GPU stack\footnote{We stress that the GPU stack is software code running on \textit{CPU}, not \textit{GPU}} has a large codebase. 
%For instance, 
Arm Mali, reported to be the most pervasive GPUs in the world~\cite{arm-mali-all}, has a runtime of a 48-MB executable; 
the driver has 45K SLoC \cite{bifrost-driver}. 
The stack often has substantial proprietary code and undocumented interfaces. 

% ML apps suffer from drawbacks of the GPU stack. 
Such a sophisticated GPU stack has created a number of difficulties. 
(1) Weak security~\cite{lee14sp,cudaLeak,sugar}.
In the year of 2020, 46 CVEs on GPU stacks were reported, most of which are attributed to the stack's complex internals and interfaces. 
% Proprietary code and interf6aces further complicate verification. 
% The complexity also creates obstacles towards porting GPU stacks in trusted execution environments, such as Arm TrustZone. 
(2) Deployment difficulty~\cite{mlFacebook}. 
%A fragmented ecosystem. 
%ML apps choose their GPU abstractions, %(e.g. Tensorflow atop OpenCL/CUDA, ncnn atop Vulkan, etc), which may only be supported on certain OSes. 
For instance, ncnn, a popular mobile ML framework, requires the Vulkan API. 
Yet the Vulkan runtime for Arm GPUs only exists on Android but not GNU/Linux or Windows~\cite{mali-vulkan-sdk}. 
% ncnn can only work with the Arm GPUs on Android but not GNU/Linux or Windows, 
% because Arm chooses to maintain their Vulkan runtimes for Android only. 
%because Arm chooses to maintain their Vulkan runtimes for Android only. 
% Vulkan for Mali GPUs is available on Android but not on GNU/Linux or Windows. 
Even on a supported OS, 
% it is not uncommon that 
an ML app often only works with specific combinations of runtime/kernel versions~\cite{tf-gpus,khronos-conformant,cuda-compat}. 
% (3) Launch overhead. 
(3) Slow startup. 
Even a simple GPU job may take several seconds to launch because of expensive stack initialization. 
This paper will show more details. 
% \sect{eval} will show more details. 
% A single API call, e.g. CLXXX, goes through many layers, first by the the runtime and then dispatched by the OS kernel.
% A microbenchmark by us shows that the XXX ms ....

%For instance, 
%Vulkan API for Mali GPUs is only available on Android but not Linux; 
%OpenCL for Broadcom GPUs is available on Linux but not Windows/Android;
%OpenCL for Adreno is available on Android but not Linux.

%%%%%%%%%%%%%%%%%%%%%%%%%%%%%%%%%%%%%%%%%%%%%%%%%%%%%%%%%%%%%%%%%%%%%%%%%%%%%%%%%%%%%%%%%%%%%%%%%%%%%%%%
%%%%%%%%%% pitfalls in using GPU stack  %%%%%%%%%%
%%%%%%%%%%%%%%%%%%%%%%%%%%%%%%%%%%%%%%%%%%%%%%%%%%%%%%%%%%%%%%%%%%%%%%%%%%%%%%%%%%%%%%%%%%%%%%%%%%%%%%%%

% !TeX root = main.tex

% this goes to a table if needed
% The stack contains user-space runtime (143 MB for userspace runtime of Mali GPU) with complex dependencies and device driver (e.g. 52 KSLoC of Mali GPU driver);

%%%%%%%%%%%%%%%%%%%%%%%%%%%%%%%%%%%%%%%%%%%%%%%%%%%%%%%%%%%%%%%%%%%%%%%%%%%%%%%%
%%%%%%%%%% Opportunities with client-side compute  %%%%%%%%%%
%%%%%%%%%%%%%%%%%%%%%%%%%%%%%%%%%%%%%%%%%%%%%%%%%%%%%%%%%%%%%%%%%%%%%%%%%%%%%%%%

%The root reason is that a GPU stack was architected for graphics at the beginning; 
The complexity of a GPU stack was mostly for its original design goal: to support \textit{interactive} apps with numerous dynamic GPU jobs. 
% its complexity is primarily for the \textit{dynamism} of interactive apps, such as GUIs and and games. 
% its complexity is primarily for \textit{dynamism} in interactive apps, which generate numerous and dynamic GPU jobs. 
Such a goal is less important to ML apps, which often run a \textit{prescribed} set of GPU jobs (albeit on new input data)~\cite{mengwei19www}; 
many ML apps run GPU job batches without user interactions; 
they can multiplex on GPU at long intervals, e.g. seconds. 
% Compiling and optimizing shaders ahead of time, instead of on target devices, becomes a norm~\cite{tvm}.
% (2) Many ML workloads, e.g. on-device training, run GPU compute in large batches without users involved.  
% They tolerate longer delays (seconds) and coarse-grained GPU sharing. 
%Each GPU job last much longer than their graphics counterparts. 
% Therefore, while ML apps still need the GPU stack for 
% While the complex GPU stack is still indispensable in translating programming abstractions (e.g. OpenCL) to GPU primitives, most of the work can be done statically. 
%Therefore, the ML apps should not be burdened by a full GPU stack. 
%They just need a fast path to push computations to GPU. 
The ML apps just need to quickly shove computation into GPU. 
They should not be burdened by a full-blown GPU stack. 
% They should be relieved from the burden of a GPU stack. 
% They should not pay the overhead of a full GPU stack. 

%%%%%%%%%%%%%%%%%%%%%%%%%%%%%%%%%%%%%%%%%%%%%%%%%%%%%%%%%%%%%%%%%%%%%%%%%%%%%%%%%%
%%%%%%%%%% Proposal - GPU activities replay  %%%%%%%%%%
%%%%%%%%%%%%%%%%%%%%%%%%%%%%%%%%%%%%%%%%%%%%%%%%%%%%%%%%%%%%%%%%%%%%%%%%%%%%%%%%%%

%\paragraph{Approach: GPU record \& replay}
\paragraph{Our approach}
\sysfull{} (\sys{}) is a new way to deploy and execute GPU compute with little changes to the existing GPU stack. 
% without overhauling the existing GPU stack. 
We focus on integrated GPUs on system-on-chips (SoCs).
%To relieve client ML apps from a full GPU stack, we propose \sys{}. 
Figure~\ref{fig:overview} overviews its workflow. 
At development time, developers run their ML app and record GPU executions. % and include the recording as part of the app. 
% This essentially \textit{flattens} the GPU stack at run time.
% Recording GPU execution is possible: 
The recording is feasible: 
despite much of the GPU stack is a blackbox,
it interacts with the GPU at a narrow interface -- registers and memory, which is managed by an open-source driver. % mapped via the GPU pagetables. 
% The interface is managed by a GPU driver, which is open-source for most GPUs. 
Through lightweight instrumentation, 
an in-driver recorder can trace CPU/GPU interactions as a series of register accesses and memory dumps which enclose proprietary GPU commands and instructions.
They are sufficient for reproducing the GPU computation.

To replay, an ML app invokes the recorded GPU executions on new input data. 
To the app, the GPU stack is substituted by a replayer, 
which is much simpler as it avoids GPU API translation, code generation, and resource management. 
% does not need to translate APIs, generate GPU shaders, manage CPU/GPU resources, or multiplex multiple apps. 
It simply accesses GPU registers and loads memory dumps at specified time intervals. 
%The replayer remain oblivious to the semantics of replayed actions. 
%No reverse engineering is required. 
Throughout the process, 
the recorder/replayer remain oblivious to the semantics of most register accesses and memory dumps. % treating them as a blackbox. 
% No reverse engineering is needed. 

\paragraph{Use cases}
%Figure~\ref{fig:deployment} show three use cases of the \sys{} replayer. 
Figure~\ref{fig:overview} shows deployment scenarios of the replayer. 

\begin{myitemize}		
	\item [D1. ]
	\textit{Co-existing with a GPU stack on the same OS.} 
	This applies to smartphones. 
	% where a mix of interactive and non- ML apps share the GPU, 
%	The interactive apps (including their ML compute) run on the GPU stack as usual. 
	Common interactive apps without \sys{} run on the GPU stack. 
%	When they are not using GPU, the OS deactivates the GPU stack and runs non-interactive ML workloads with replay. 
	When they are not using GPU, the OS runs \sys{}-supported ML with replay. 
	% In case of impromptu GPU needs by the interactive apps, 
	Once the interactive apps ask for GPU, 
	the OS preempts GPU from the ongoing replay with short delays (\sect{replayer}). 
%	\sys{} benefits the non-interactive ML with deployment ease and faster GPU invocations.
%	The ML workloads enjoy deployment ease and faster GPU invocations.
	 							
	\item [D2. ]
	\textit{In TEE.}
	This applies to Arm TrustZone~\cite{streambox-tz}. 
	On the same machine, apps not using \sys{} run on the GPU stack in the normal world and \sys{}-supported ML runs atop a replayer in the secure world. 
%	The secure world code invokes the replayer, e.g. for analyzing privacy-sensitive data~\cite{HIX,streambox-tz}.
	A secure monitor at EL3 switches GPU between the two worlds. 
% 	Being minimalistic and standalone, the replayer can be instantiated in the secure world with low effort and a low chance of vulnerabilities. 
%	For instance, the monitor allocates sufficient GPU time to TEE; 	it makes the TEE yield GPU to short, bursty graphics jobs outside the TEE. 

%	To support SGX is not as simple as TrustZone due to its ban on MMIO access from enclave;
%	\sys{} can be applied to SGX with MMIO remoting or extending SGX instructions as HIX proposed~\cite{HIX}.

%	Note that \sys{} does not support vanilla SGX, which bans direct MMIO access from enclave.
%	it can be extended to SGX with MMIO remoting or support of additional SGX instructions as HIX did~\cite{HIX}.

	\item [D3. ]
	\textit{As a replacement for the system's GPU stack.} 
	This applies to headless devices such as robots, 
	where \sys{}-supported ML apps share GPU cooperatively. 
	Each ML app runs its own replayer instance. % and yield to each other. %  at the ends of recordings as orchestrated by the OS.
	%  one app finishes a recording before yielding GPU to the next app. 
\end{myitemize}

\paragraph{Benefits}
\sys{} offers the following benefits:

%\noindent
(1) \textit{Security} \hspace{1mm}
%First, \sys{} guards the attack surface of ML apps: 
%they no longer rely on a GPU stack on the client machines but the stack 
%on the developer's machines for recording.
First,
\sys{} better shields the GPU stack. 
% the GPU stack serving ML apps no longer runs on the client machines but on the developer's machine.  
%The GPU stack serving ML apps is decoupled from target machines and is instead located on the developer's machine.  
The GPU stack serving the target ML app is detached from the app and instead resides on the developer's machine for recording only.
Hence, the stack is no longer exposed to many threats in the wild but instead 
%~\cite{CVE-2014-1376,CVE-2019-5068,CVE-2018-6253,CVE-2017-18643,CVE-2019-20577,CVE-2020-11179,CVE-2019-10520,CVE-2014-0972,CVE-2019-14615}; 
protected as part of software supplychain, for which attacks require high capabilities and long commitment~\cite{nist-supplychain}.
% Second, \sys{} replaces the GPU stack on the target machines with the replayer with fewer vulnerabilities. 
Second, on target machines, the replayer replaces the GPU stack for the app (D1/D2) or for the whole system (D3). 
As a result, either the app or the whole system is free from vulnerabilities from the GPU stack,
which originate in rich features such as buffer management~\cite{CVE-2019-20577,CVE-2014-0972} and fine-grained sharing~\cite{CVE-2017-18643,CVE-2020-11179,CVE-2019-14615},
as well as complex interfaces such as framework APIs~\cite{CVE-2014-1376},  IOCTLs~\cite{CVE-2019-10520}, and directly mapped memory~\cite{CVE-2019-5068}. 
By comparison, 
% the replayer does not have such features. 
the replayer only has a few K SLoC and exposes several simple functions; 
replay actions have simple, well-defined semantics and are amenable to checks. 
(2) \textit{Ease of ML deployment} \hspace{1mm}
The replayer can run in various environments: at user or kernel level of a commodity OS, in a TEE, in a library OS, and even baremetal. 
Section~\ref{sec:impl} will present the details. 
\sys{} brings mature GPU compute such as Tensorflow NNs to these environments without porting full GPU stacks. 
%Section~\ref{sec:impl} presents various replayer implementations.
% We will report to run tensorflow models on baremetal hardware. (Sec XXX)
%\sys{} works for integrated GPUs from multiple vendors and 
\sys{} is compatible with today's GPU ecosystems. 
It requires no reverse engineering of proprietary GPU runtimes, commands, and shaders. 
Agnostic to GPU APIs, \sys{} can record and replay diverse ML workloads.
%It is agnostic to GPU APIs and thus supports diverse ML workloads. 
%Agnostic to programming abstractions, \sys{} can record and replay diverse ML workloads.
%\sys{} can record and replay diverse ML workloads. 

\begin{comment}
%\noindent
(3) \textit{Compatible with GPU ecosystems} \hspace{1mm}
We show that \sys{} works for integrated GPUs from multiple vendors. 
It requires no reverse engineering of proprietary GPU runtimes, GPU commands, and shader instructions. 
Agnostic to programming abstractions, 
\sys{} can record and replay diverse ML workloads. 
% OpenCL, Vulkan, and GLES compute. 
\end{comment}

%(iv) 
%\textit{Faster GPU invocation} --
%\sys{} reduces the CPU-side overhead to baremetal: register accesses and copying of GPU memory. 
%It removes expensive abstractions of multiple software layers, dynamic CPU/GPU memory management, and just-in-time generations GPU commands and code. 

(3) 
\textit{Faster GPU invocation} \hspace{1mm}
\sys{} reduces the GPU stack initialization to baremetal: register accesses and GPU memory copy. 
It removes expensive abstractions of multiple software layers, dynamic CPU/GPU memory management, and just-in-time generation of GPU commands and code.

%%%%%%%%%%%%%%%%%%%%%%%%%%%%%%%%%%%%%%%%%%%%%%%%%%%%%%%%%%%%%%%%%%%%%%%%%%%%%%%%%
%%%%%%%%%% Challenges  %%%%%%%%%%
%%%%%%%%%%%%%%%%%%%%%%%%%%%%%%%%%%%%%%%%%%%%%%%%%%%%%%%%%%%%%%%%%%%%%%%%%%%%%%%%%%

%\paragraph{Challenges:}
%The above approach is challenged by the ... ii) Nondeterminism in GPU execution that could fail the replay; iii) handling recorded delays (?); iv) recorder/replayer simplicity
%
%\begin{itemize}
%	\item Nondeterminism in GPU execution. We observe that the original GPU stack involves dynamic memory allocation and nondeterministic soft and hardware communication. 
%	Notably, the timing of GPU executions may vary; addresses of GPU memory objects may differ. 
%	... \Note{this could lead to replay failures.} Because the GPU may not be in exactly the same state. For instance, replaying a GPU command when it has no resource to accept the command; 
%
%	\item Handling timing/delays. Even with deterministic sequence of interactions, the timing will vary across runs. Some timing/delays (artificial) must be preserved. For instance: replay GPU commands overspeed may lead to failure. Some shall be removed for speed. 
%
%\end{itemize}	

%%%%%%%%%%%%%%%%%%%%%%%%%%%%%%%%%%%%%%%%%%%%%%%%%%%%%%%%%%%%%%%%%%%%%%%%%%%%%%%%%%
%%%%%%%%%% Contributions  %%%%%%%%%%
%%%%%%%%%%%%%%%%%%%%%%%%%%%%%%%%%%%%%%%%%%%%%%%%%%%%%%%%%%%%%%%%%%%%%%%%%%%%%%%%%%

\paragraph{Challenges}
%We have addressed several challenges. 
First, we make reproduction of GPU workloads feasible despite the GPU's complex interfaces and proprietary internals. 
We identify and capture key CPU/GPU interactions and memory states; 
we selectively dump memory regions and discover the input/output addresses operated by GPU commands/shaders.  

Second, we ensure \sys{}'s replay is correct in the face of nondeterministic CPU/GPU interactions. 
%A key insight is that replay correctness is equivalent to GPU finishing the same sequence of jobs as recorded, which is in turn equivalent to the GPU following the same path of state transitions as recorded. 
A key insight is that replay correctness is equivalent to the GPU finishing the same sequence of state transitions as recorded. 
To this end, 
we \textit{prevent} many state divergences by eliminating their sources at the record time; 
we \textit{tolerate} non-deterministic interactions that do not affect the GPU state at the replay time.
%we \textit{detect} non-preventable divergence and recover from them at the replay time.
\sys{}'s approach to nondeterminism sets it apart from prior record-and-replay systems~\cite{sanity,rtag,replayConfusion}: targeting program debugging, they seek to reproduce the original executions with high fidelity and preserve all nondeterministic events in replay.

% Third, we investigate how usable \sys{} is. 
% Third, we investigate \sys{}'s applicability and practicality. 
% Third, we investigate issues that make \sys{} practical. 
% Third, we investigate issues that make \sys{} practical. 
Third, we investigate a variety of practicality issues. 
We identify the minimum GPU hardware requirements. 
We show that \sys{} requires low developer efforts, and such efforts are often amortized over a family of GPUs supported by one driver. 
% invested once for a family of GPUs supported in one driver. 
% and study if they apply to mainstream integrated GPUs. 
We explore \sys{}'s deployment ranging from smartphones to headless IoT devices. 
% including smartphones, headless IoT devices, and TEE. 
%We investigate how to map a ML workload to \sys{} recordings and the impact of mapping granularities. 
We investigate how to map an ML workload to \sys{} recordings and quantify the impact of recording granularities. 
%We explore GPU hardware requirements for record and replay. 
%We propose the preemption mechanism for the replayer to share GPU with other GPU interactive apps in case they coexist. 
We propose a scheduling mechanism for the replayer to share GPU with interactive apps. 

\paragraph{Results}
\sys{} works on a variety of GPUs (Arm Mali and Broadcom v3d), APIs (OpenCL, GLES compute, and Vulkan), ML frameworks (ACL~\cite{acl}, ncnn~\cite{ncnn}, Tensorflow~\cite{tensorFlow}, and DeepCL~\cite{deepCL}), and 33 NN implementations.
We build replayers for userspace, kernel, TrustZone, and a baremetal environment. 
We show that 
a recording with light patching can be replayed on different GPU hardware of the same family. 
%	A recording is as small as several MBs. 
%	A recording is no larger than a few MBs. 
%	We evaluate \sys{} on inference and training of 33 NN implementations.
Compared to the original GPU stack, 
the replayer's startup delays are lower by up to two orders of magnitude; 
its execution delays range from 68\% lower to 15\% higher. 
	
This paper makes the following contributions:

\begin{myenumerate}

	\item
	\sysfull{} (\sys{}), a new way to deploy GPU computation. 
%	\sys{} builds on minimum assumptions of GPU hardware and simple knowledge on GPU interfaces. 
%	It applies to a variety of integrated GPUs. 
	
	\item 
	%A suite of techniques for identifying and recording the minimum GPU execution for replay: 
	A recorder that captures the essential GPU memory states and interactions for replay.  
	
%	%and produces recordings for robust, efficient replay. 	
%	Without requiring intimate knowledge of GPU, 
%	the recorder summarizes GPU register accesses, selectively dumps memory regions, 
%	discovers the input/output addresses operated by GPU commands/shaders, 
%	and compresses observed time intervals for fast replay. % in a recording. 
	% We explore the tradeoffs in recoding granularities. 

	\item A safe, robust replayer that verifies recordings for security, supports GPU handoff and preemption, and detects and recovers from replay failures. 
	
%	The replayer statically verifies recordings for security, 
%	including legal memory and register access; 
%	it supports GPU handoff and preemption, 
%	allowing the replayer to share GPU with other GPU apps in a cooperative fashion;  
%	%which multiplex the	replay activities and other apps on the same GPU in a 	
%	%allowing the	replay activities to share the GPU with other apps 
%	it detects and recovers from replay failures. 

	\item Realization of the design in diverse software/hardware environments. 
	
%	\sys{} works on a variety of GPUs (Arm Mali and Broadcom v3d), APIs (OpenCL, GLES compute, and Vulkan), ML frameworks (ACL~\cite{acl}, ncnn~\cite{ncnn}, Tensorflow~\cite{tensorFlow}, and DeepCL~\cite{deepCL}), and 33 NN implementations.
%	We build replayers for userspace, kernel, TrustZone, and a baremetal environment. 
%	We show that 
%	a recording with light patching can be replayed on different GPU hardware of the same family. 
%%	A recording is as small as several MBs. 
%%	A recording is no larger than a few MBs. 
%%	We evaluate \sys{} on inference and training of 33 NN implementations.
%	Compared to the original GPU stack, 
%	the replayer's startup delays are lower by up to two orders of magnitude; 
%	its execution delays range from 40\% lower to 13\% higher.
%%	the replay reduces startup delays by up to two orders of magnitude and incurs XX\%-XX\% lower delay than the original GPU executions. 

\end{myenumerate}

\section{Motivations}
\label{sec:bkgnd}

\subsection{The GPU Stack and Its Problems}
\label{sec:gpu-stack}

% !TeX root = main.tex

\begin{figure}[t!]
\centering
\includegraphics[width=0.43\textwidth{}]{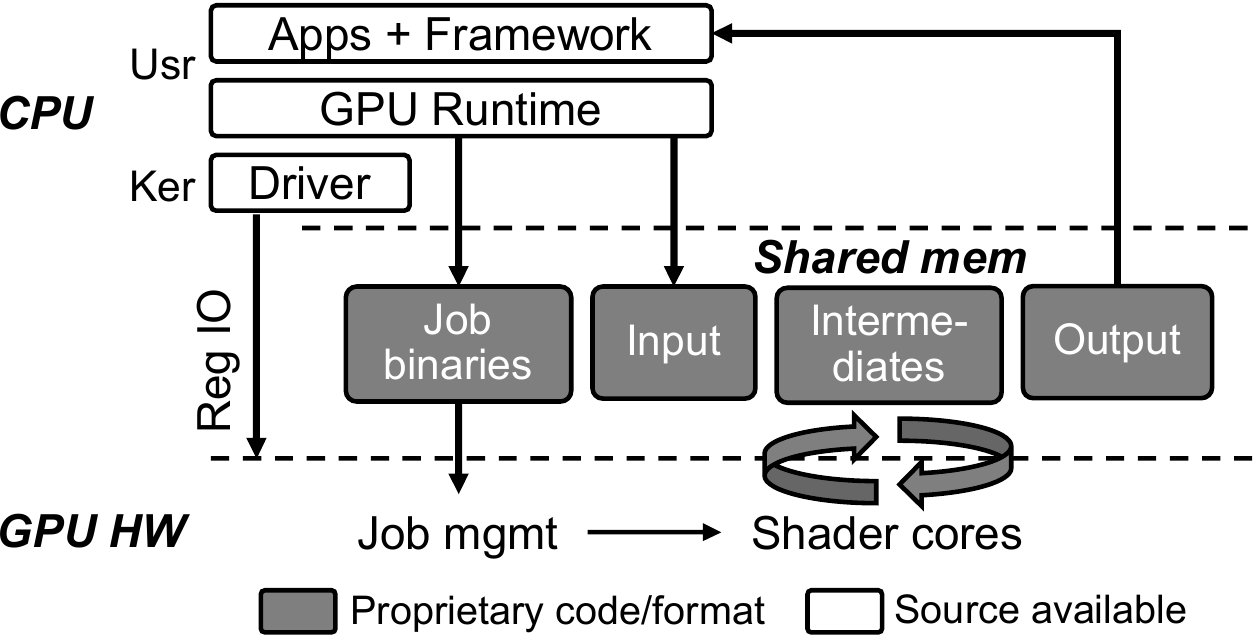} 
%\missingfigure[figwidth=6cm]{Juno platform power instrumentation}
%\vspace{-8pt}		% use as needed
\caption{The software/hardware for an integrated GPU}
\label{fig:job_exec}
\label{fig:arch}
\vspace{-8pt}		% use as needed
\end{figure}

% Yet, the graphics complexity becomes an overkill for compute. 

\paragraph{CPU/GPU interactions}
As shown in Figure~\ref{fig:job_exec}, 
CPUs request computation on GPUs by sending jobs to the latter. 
The GPU runtime directly emits GPU job binaries -- GPU commands, metadata, and shaders -- to GPU-visible memory\footnote{GPU memory for short, with the understanding it is part of shared DRAM}. 
The runtime communicates with the driver with ioctl syscalls, e.g. to allocate GPU memory or to start a job. 	
%Of a typical GPU stack, the runtime is released by vendors as binaries;  the GPU job formats and data layout are proprietary and undocumented;  the driver is often open source. 

% \input{fig-nondeterminism}
%\paragraph{Nondeterminism on GPU execution}

%\paragraph{Features for dynamism}
\paragraph{Why are GPU stacks complex?}
% While today's GPU stack is dual modality for both graphics and compute, 
% its key features cater to graphics. 
Several key features of a GPU stack cater to graphics. 

%For instance, memory management; texture management; conversion. Entangled. Vulkan, GL compute, ..., even OpenGL/CUDA share the same kernel driver with graphics. 

%\paragraph{The design for graphics}
%Several top features of a GPU stack cater to graphics. 

%These apps render visual effects based on user inputs, 
%necessitating online generation of GPU commands and executable codes; 
%for low delays, these apps stream small GPU job batches; 
%multiple apps often render with GPUs concurrently, calling for fine-grained, dynamic sharing (multiple contexts, dynamic memory) of the GPU hardware. 

\begin{myenumerate}						
	\item \textit{Just-in-time (JIT) job generation.}	
	Graphics apps emit numerous GPU jobs, from uploading textures to rendering fragments. 
	% -- uploading textures, calculating vertices, rendering fragments, etc. 
		For instance, during a game demo of 50 seconds~\cite{q3-demo}, the v3d GPU executes 32K jobs.  % each lasting 1.5 ms on average. 
		A game may rewrite shader sources for jobs~\cite{doom3-convert-shader}.
	Unable to foresee these jobs,
	the GPU stack generates their commands and shaders just in time. 
			
	\item \textit{Dynamic resource management.}
% 	Driven by user interactions, 
	Depending on user interactions, 
	graphics apps generate GPU jobs with various input sizes, data formats, and buffer lengths. 
	% e.g. depending on UI layout or user's navigation in a game scene. 
	They require dynamic management of GPU time and memory, which may further entail sophisticated CPU/GPU coordination~\cite{dissect-apple-m1}.
%	CPU and GPU may collaborate to manage memory via a proprietary coordination protocol~\cite{dissect-apple-m1}. 
	%on some GPUs, this further requires sophisticated CPU/GPU collaboration~\cite{dissect-apple-m1}

	\item \textit{Fine-grained multiplexing}.
%	On desktops or smartphones,
	Concurrent programs may draw on their screen regions. 
	To support them, the GPU stack interleaves jobs at fine  intervals and maintains separation. 
%	e.g. by swapping GPU page tables among these programs. 
\end{myenumerate}
		
%	\item{Multiple contexts (multiple GUI windows)}: The user can launch multiple GPU application at a time that results in multiple GPU contexts; even a single application has multiple GPU contexts. These contexts share a single GPU which necessitates GPU context scheduling from the GPU stack. For instance, a window manager can create more contexts when user run a new application within a new window, which requires graphics rendering. 
%%	Note that the multiple GPU contexts may result in unpredictable impact of cache miss and flush.
%
%	\item{Fine-grained sharing \jin{can be merged with the above}}:serving multiple apps which submit (small) jobs at fine temporal intervals; 
%	recent GPUs support simultaneous execution of multiple apps/contexts (spatial multiplexing)\jin{evidence?}, 
%	each using a subset of GPU cores. 
%	the graphics workload may involve a number of GPU contexts; the device driver picks up the next context based on its scheduling policy.  
	
%	\item{JIT compilation}:
%	The prescribed/general computation only uses a set of kernels within pre-defined pipeline, which allows ahead of time compilation. However, the graphics workload is dynamic and unpredictable and thus requires JIT compilation when a new kernel need to be added (e.g. when user runs a new application while there was a running GPU application).

%\paragraph{Compute workloads} Compute for ML shows disparate nature.

\paragraph{Compute for ML} shows disparate nature unlike graphics.

\textit{Prescribed GPU jobs}:
One app often runs pre-defined ML algorithms~\cite{mengwei19www}, requesting a smaller set of GPU jobs repeatedly executed on different inputs. 
% Popular mobile/edge frameworks for neural networks (NN), e.g. ARM ACL~\cite{acl} or Tencent ncnn~\cite{ncnn}, ship with only a few hundreds of distinct shaders in total. 
Popular neural networks (NN) often have tens of GPU jobs each (\S\ref{sec:eval}). 
The needed GPU memory and time can be statically determined.
%when the ML models were designed.
% GPU jobs for ML tend to last much longer. 
% For instance, executing AlexNet for inference on Arm Mali only runs 45 distinct GPU jobs.
% As show in Figure XXX) each job in AlexNet lasts XXX on average, as opposed to XXX on a game demo. 

%	Unlike graphics workloads that generate a large number of GPU jobs on the fly (\S~\ref{sec:gpu-stack}), 
%	GPU often runs compute kernels that have code and parameters (e.g. input/output sizes) are pre-determined at development time, e.g. by NN designers. 
%	The number of \textit{distinct} compute kernels is low, despite that these kernels may run repeatedly, e.g. on different inputs. 
%	Figure XXX shows that, 
%	AlexNet and MobileNet, two popular NNs on client devices, have XXX and XXX distinct compute kernels, respectively. 
%	Popular client ML frameworks, e.g. ARM ACL or Tencent ncnn, implement XXX distinct kernels. 
%	A smartphone app often runs a small set of pre-defined NN networks.  etc)~\cite{mengwei19www}.

\textit{Coarse-grained multiplexing}:
On embedded devices, ML may run on GPU for long without sharing (e.g. object detection on a smart camera). 
On multiprogrammed smartphones, ML apps may run in background, e.g. model fine-tuning. 
% e.g. photo beautification or model retraining. 
Such an app tolerates delays of hundreds of milliseconds or seconds in waiting for a GPU; once on GPU, it can generate adequate workloads to utilize the GPU. 
%These apps can be served with simple multiplexing of GPU. % and GPU handoff mechanisms. 

%\paragraph{Coping with software blackboxes}
%\paragraph{Respect software blackboxes}
%\paragraph{Respect runtime blackboxes \& exploit opensource drivers}
\paragraph{Runtime blackboxes} 
Most GPUs have proprietary runtime, job binaries, and shaders. 
While \sys{} can be more efficient had it known these internals or changed them, 
% e.g. to selectively capture GPU memory, 
doing so requires deep reverse engineering and makes deployment harder. 
%defeats our goal of engineering ease. 
% and results in GPU-specific designs. We hence assume runtime and jobs are blackboxes.
Hence, we avoid changing these blackboxes but only tap in the Linux GPU drivers which are required to be open-source. 
% requiring deep knowledge makes \sys{} less generic (GPUs can be vastly different) and less practical (some knowledge may be undisclosed and require reverse-engineering). 
%On the other hand, 
%
%\sys{} taps in the Linux GPU drivers which are required to be open-source. We examine the driver code to learn GPU interfaces and build the recorder by adding hooks to the driver. 
%This effort is modest and applies to most integrated GPUs we know.
%Section~\ref{sec:abs_model} presents a case study. 
% Table~\ref{sec:} will present case studies. 
% The only few exceptions include Apple M1 and XXX, for which reverse-engineering to obtain the above information would be required.  

\paragraph{Design Implication}
A GPU stack's dual modality for graphics and compute becomes a burden. 
While an ML app still needs the GPU stack for translating higher-level programming abstractions to GPU hardware operations, 
the translation can happen ahead of deployment. 
At run time, the ML app just needs a simple path to push the resultant operations to GPU.

%partitioning the stack \st{for compute and graphics} is infeasible. 
%(1) for compute, the stack is still indispensable in translating programming abstractions (e.g. OpenCL) to GPU primitives; 
%(2) Re-engineering the stack -- from the runtime to device driver -- and maintaining the separation, requires formidable efforts. 
%\Note{list some evidence. SLOCCOUNT, etc.}
%Quite the opposite, recent GPU development favors tighter graphics/compute integration. 
%For instance, Vulkan, a GPU abstraction that supports both graphics and compute, sees far wider adoption by mobile devices than the compute-only OpenCL \Note{cite}. 

% much of the work can be done ahead of time. 
% it demands very lightweight system support. 
% it can run atop lightweight system support. 
% They deserve a far more lightweight stack. 

%\subsection{Embedded GPU Trends}
%\subsection{GPU Trends for Compute}
\subsection{GPU Trends We Exploit}
\label{sec:bkgnd:trend}
\label{sec:gpu-trend}

% In embracing compute, modern GPUs show the following trends that make \sys{} possible.

%The following hardware trends make \sys{} possible.

% \paragraph{GPUs run on virtual memory}
\paragraph{GPU virtual memory}
Today, most integrated GPUs run on virtual address spaces. 
% To configure a GPU's address space, the GPU stack populates the GPU's page tables and sets an MMU register of GPU's to the page table base. The runtime links GPU commands and shaders to virtual addresses. 
To configure a GPU's address space, the GPU stack populates the GPU's page tables and links GPU commands and shaders to the virtual addresses. 
% GPU commands and shaders are linked to virtual addresses. 
% The CPU and the GPU may map a shared physical memory region to different addresses in their respective spaces. 

% The GPU MMU is activated by the device driver as follows: it 1) allocates GPU memory as needed and maps them into CPU and/or GPU address space, 2) constructs a distinct page table for the GPU, which reflects all the GPU to host memory mappings, 3) activates the page table on the GPU MMU by synchronizing modified entries or updating PGD directly. Note that the page table can be updated either when the beginning of GPU computation or even during run time depending on the implementation of application using GPU.

%\paragraph{GPUs are more autonomous}
\paragraph{GPU autonomy}
% A job consists of a sequence of GPU commands (e.g. for moving data and converting format) and one or more shaders. 
%Modern GPU hardware is increasingly autonomous. 
%To this end, 
To reduce CPU overhead, 
a GPU job packs in much complexity -- control flows, data dependency, and core schedule. % To configure a GPU's address space, the GPU stack populates the GPU's page tables and sets an MMU register of GPU's to the page table base. The runtime links GPU commands and shaders to virtual addresses. . 
The GPU parses a job's binary, resolves dependency, and dispatches compute to shader cores. 
% It only needs to notify the CPU when the job is done. 
A job may run as long as a few seconds without CPU intervention.

Take Mali G71 as an example: 
a job (called a ``job chain'') encloses multiple sub jobs and the dependencies of sub jobs as a chain. 
To run AlexNet for inference, 
the runtime (ACL v20.05) submits 45 GPU jobs, 
5--6 GPU jobs per NN layer; 
the GPU hardware schedules a job over 8 shader cores. 

% \Note{dependency management is key. and GPU side scheduling}

% !TeX root = main.tex

\begin{figure}
	\centering
	\includegraphics[width=0.45\textwidth{}]{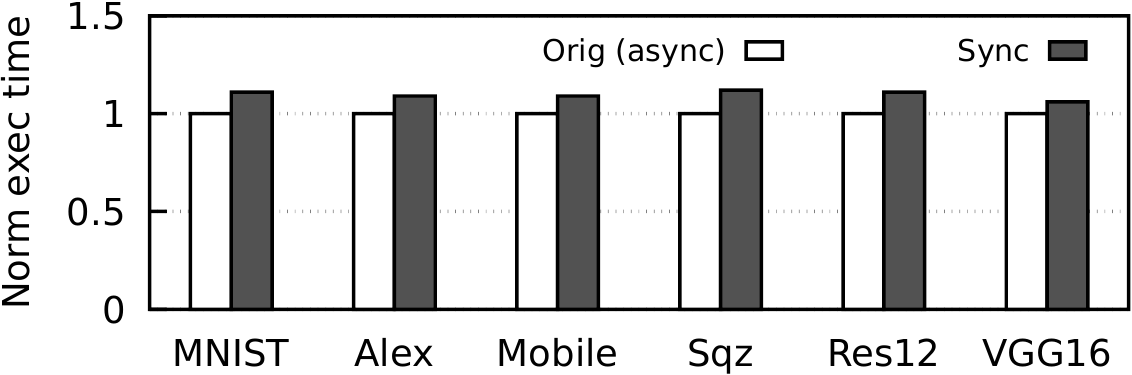} 
	%\missingfigure[figwidth=6cm]{Juno platform power instrumentation}
%	\vspace{-10pt}		% use as needed
	\caption{Synchronous job submission increases NN inference delay modestly. ACL~\cite{acl} + OpenCL on Mali G71}
	%\caption{The nondeterministic behavior of GPU and its software stack interaction}
	\label{fig:sync-overhead}
	\vspace{-8pt}		% use as needed
\end{figure}

\paragraph{Synchronous job submission}	
%Asynchronous GPU job submission used to be the norm: 
%the GPU executes a stream of small jobs for graphics and interrupts CPU frequently for assistance. 
Asynchronous GPU job submission is crucial to graphics, 
for which GPU executes smaller jobs. % and interrupts CPU frequently. 
% 	For instance, a game demo (Quake3 on Broadcom VideoCore6) submits XXX GPU jobs, each lasting XXX ms on average. 
To hide job management delays, 
%(preparation, submission, and cleanup),
CPU streams jobs to GPU to keep the latter busy. 
% CPU and GPU produce and consume jobs in a streaming fashion to stay busy simultaneously. 
% shared a deep job queue (e.g. 64 slots for XXX??),
% For large compute jobs, asynchronous submission is less important: 
% For compute, however, a job's management delay (sub millisecond) is amortized over the job's longer execution (often tens of milliseconds or longer). 
Yet for compute, a job's management delay is amortized over the job's longer execution. % (often tens of milliseconds or longer). 
For simplicity, shallow job queues in GPU drivers are common (max two outstanding jobs in Mali~\cite{mali-submit} and one in  v3d/vc4~\cite{v3d-driver,vc4-submitcl}).
Figure~\ref{fig:sync-overhead} shows that synchronous job submissions incur minor computation performance overhead: 
%
% We confirm the low overhead of synchronous jobs:
with six NN inferences on Mali G71 (see Table~\ref{tab:model_desc} for details), we find that enforcing synchronous jobs only adds 4\% delays on average (max: 11\%, min: 2\%).

%and toggle asynchronous job submission in the GPU driver.
%The result shows that synchronous job submission only adds 4\% delays on average (max: 11\%, min: 2\%).

%Our experiments confirm the low overhead of synchronous job submission. 
%We run inferences with six NNs on Mali G71 (see Table~\ref{tab:model_desc} for details) and toggle asynchronous job submission in the GPU driver. 
%Our measurement shows that synchronous job submission only adds 4\% delays on average (max: 11\%, min: 2\%).

\subsection{Design Choices}
The trends above motivate the following choices. 

\sys{} focuses on synchronous GPU jobs, queuing them and executing one job at a time. 
It eschews recording or replaying concurrent GPU jobs. 
This deliberate decision ensures replay determinism: 
with concurrent GPU jobs, the number of possible CPU/GPU interactions would grow exponentially, making faithful replay difficult.
The overhead of synchronous jobs is low as shown above.

For the same reason, \sys{} eschews GPU sharing across apps during record and replay. 
Even without sharing, \sys{} has important use cases. 
On smartphones, examples include background ML such as photo beautification and model fine-tuning; on headless smart devices without graphics, examples include ML pipelines for vision and prediction.  
Furthermore, the replayer can yield GPU to interactive apps with low delays (\S\ref{sec:replayer}).

%\paragraph{The boundary}
% This leads to a small trusted computing base (TCB) for GPU computation. 
% With little external dependency, the replayer can be ported to almost any runtime environments. 
\sys{} records at the lowest software level, i.e. the CPU/GPU boundary. % 
This makes the replayer small and portable. 
%, which is crucial to its security and  portability. 
By contrast, recording at higher levels, e.g. GPU APIs~\cite{telekine} or ML frameworks~\cite{nimble}, 
would require the replayer to incorporate extensive runtime or driver functionalities.

\section{\sys{}}
\label{sec:overview}
% This section overviews \sys{}: its use cases, design considerations, and assumptions. 

%\subsection{Overview}

\subsection{Using \sys{}}

% !TeX root = main.tex

\begin{figure}
	\centering
	\includegraphics[width=0.45\textwidth{}]{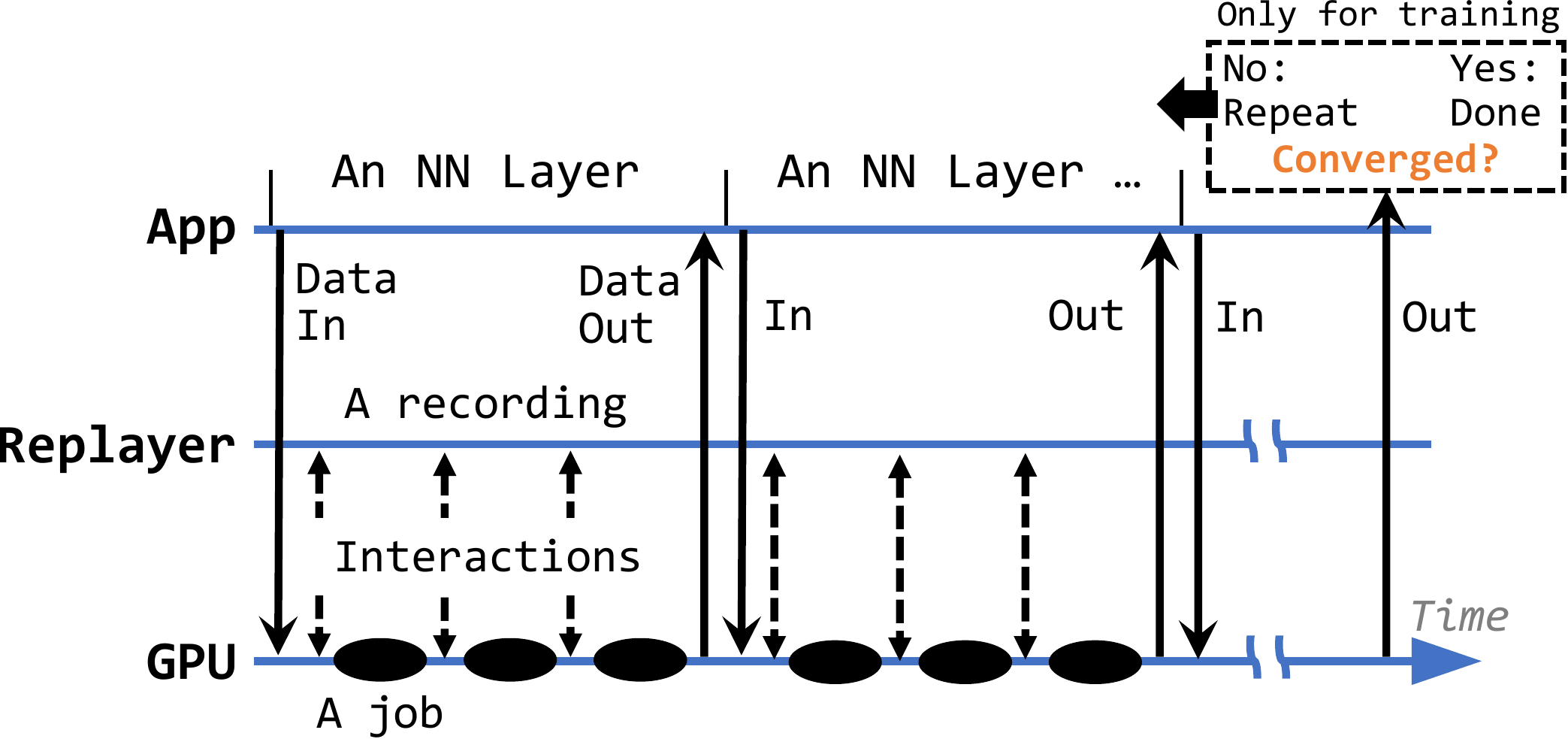} 
	%\missingfigure[figwidth=6cm]{Juno platform power instrumentation}
	\vspace{-3.5pt}		% use as needed
%	\caption{Using \sys{} for replay}
	\caption{Replaying NN execution with \sys{}}
	%\caption{The nondeterministic behavior of GPU and its software stack interaction}
	\label{fig:using}
	\vspace{-8pt}		% use as needed
\end{figure}

%\noindent
%\textbf{A recording} is a linear sequence of CPU/GPU interactions accompanied by GPU memory dumps. 
%\textbf{A recording} 
\paragraph{A recording}
encodes a fixed sequence of GPU jobs, 
including the CPU/GPU interactions 
%(in an XML file) 
and GPU memory dumps 
%(in ELF files) 
needed to execute these jobs.
%they will be delivered to a replayer as a bundle of XML (replay actions in Table~\ref{tab:ir}) and ELF files (memory dumps).
%For a recording to include all GPU jobs of a workload, 
%To ensure a workload's all GPU jobs included in a recording,  
%the workload must execute all the jobs regardless of input data, 
%the workload must execute all the jobs regardless of input data,
To capture a workload in one recording,
the workload is required to execute all its jobs regardless of input,
%For a workload to be captured in one recording, 
%the workload is required to execute all its jobs regardless of input, 
i.e. the workload's job graph contains no conditional branches that lead to different types of GPU jobs. 
The requirement does not preclude conditional branches \textit{inside} a GPU job, i.e. among GPU instructions.
This is because \sys{} dumps a job's entire binary, which includes all the branches within, no matter whether they were exercised at the record time. 
%This is because \sys{} dumps a job's entire binary, which includes all the branches within, no matter whether these branches were exercised at the record time. 

The above requirement is met by most, if not all, popular NNs, including 
all 44 NNs shipped with ACL, ncnn, and Tensorflow~\cite{acl,ncnn,tf-examples}.
% which execute all their GPU jobs unconditionally. 
%They include all XXX NNs shipped with ACL, ncnn, and Tensorflow. \Note{cite}
Note that some NNs (e.g. SqueezeNet and GoogLeNet) use ``branches'' 
to refer to routes in their job graphs, which are in fact executed unconditionally. 

As examples, Figure~\ref{fig:using} shows two common NN workloads. % NN execution. 

\begin{myitemize}
\item 
%Example 1 (NN inference): The workload 
\textit{NN inference} runs a sequence of NN layers $\{L_1...L_n\}$, each executing a sequence of GPU jobs unconditionally. 
To record, developers run the inference once and create recordings $\{R_1...R_n\}$, one recording per NN layer. 
%An ML app replays the recordings corresponding to the layers. 
An ML app supplies input and replays $\{R_1...R_n\}$ in sequence. % corresponding to the layers.
After the replay, the replayer extracts output from GPU memory to the app. 

\item 
%Example 2 (NN training): The workload 
\textit{NN training} runs a sequence of NN layers $\{L_1...L_n\}$ iteratively; after each iteration, 
it evaluates a predicate $\mathscr{P}$ and terminates if $\mathscr{P}$ shows the result has converged.
%it tests the result and terminates if the result has converged.
To record, developers run one iteration and create a sequence of recordings $\{R_1...R_n\}$.
They do not handle conditionals. 
An ML app runs a training iteration by replaying $\{R_1...R_n\}$.
After the iteration, the app code on CPU evaluates $\mathscr{P}$. 
%At the iteration end, the app code on CPU evaluates $\mathscr{P}$. 
% on the result returned by GPU; 
% It repeats replaying $\{R_1...R_n\}$ on updated input data until the result converges. 
%If the results has not converged, 
Unless $\mathscr{P}$ shows convergence, 
the app replays $\{R_1...R_n\}$ again on refined input. 

%the app replays $\{R_1...R_n\}$ again on a refined input. 

\begin{comment}
\item 
Example 2 (a variant of above): CPU launches a big GPU job, which executes a training loop internally until convergence. Developers record the big job once and replay it once. The loop may run different iterations depending on the input. Developers are oblivious, because these branches are *within* the job. 
\end{comment}
\end{myitemize}

The only exception to the above requirements, to our knowledge, is a conditional NN~\cite{DBLP:journals/corr/IoannouRZKSBC16} using branches to choose among normal NNs. % as an ensemble. 
In this case, developers record branches as separate recordings; 
at run time, an ML app evaluates branch conditions on CPU and conditionally replays recordings. 
%Conditional NNs are far less common in practice. 
Conditional NNs are rare in practice to our knowledge.

%at the run time, the target app replays recordings for the network branches conditionally executed.  
% conditionally. 

% An app developer may record her own ML workloads to be part of the app;she may also link her app to well-known GPU recordings as libraries.

% Note that a GPU job \textit{can} have conditional code internally, as \sys{} captures GPU jobs in full and hence all their internal branches. 

\paragraph{CPU/GPU coordination}
Beyond the examples above, \sys{} supports a workload consisting of interleaved CPU/GPU phases. 
For such a workload, 
the recorder generates multiple recordings, one recording per GPU phase. 
At run time, the app executes the CPU phases (not recorded) and replays for the GPU phases. 
% Example: a sequence of NN layers, some on CPU and some on GPU. The recorder produces per-layer recordings for GPU layers; the target app executes CPU layers (which GPURip does not record) and replays the GPU layers in between. 

Such a hybrid execution is possible because \sys{} stitches CPU and GPU phases by their input/output. 
To do so, the recorder automatically discovers input/output addresses for GPU recordings; before and after replaying each recording, the replayer deposits/extracts data to/from the GPU memory, respectively. 
In particular, CPU/GPU synchronizations (e.g. CPU waits for an OpenCL event) are recorded/replayed by \sys{} as waits for GPU interrupts at the driver level.
See \sect{recording} for details.

%\paragraph{Recording granularity} is a developers' choice. 
%It is a tradeoff between composability and efficiency; 
\paragraph{Recording granularity} is a tradeoff between composability and efficiency; 
it does not affect correctness. 
In the examples above, developers record separate NN layers; 
% This is feasible as NNs often build on a modest set of layer types and then the shader types~\cite{mengwei19www}.
% , e.g. convolutional layers and fully connected layers. 
% (e.g. 3 \texttt{conv} and 1 \texttt{fc} layers for MNIST, 5 \texttt{conv} and 3 \texttt{fc} layers for AlexNet; 13 \texttt{conv} and 3 \texttt{fc} layers for VGG16).
% This is feasible as a modern ML only builds on a modest set of compute shaders (XXX for XXX, XXX for XXX). 
alternatively, they may record a whole NN execution as one recording. 
While per-layer recordings allow apps to assemble new NNs programmatically, a monolithic recording improves replay efficiency due to reduction in data move and cross-job optimizations. 
%The latter choice reduces data move in replay, better amortizes the GPU initialization overhead, and encompasses cross-job optimizations in the resultant recording. 
Section~\ref{sec:eval} will evaluate these choices. 
%As a result, replay incurs less data move between the target app and GPU; 
% replay incurs less data move as one job's result can remain in GPU memory without moving back to the app; 
%the initialization overhead, e.g. GPU cache flush, is amortized over many GPU jobs; 
%the resultant recording may enclose cross-job optimizations, e.g. layer fusion~\cite{tvm}. 

\paragraph{Recording portability}
By default, \sys{} expects the GPU hardware (SKUs) and firmware versions used for record and replay to exactly match.
As \sect{impl} will show, record/replay with different SKUs of the same family is possible, yet lightweight patching is needed.

\paragraph{Developer efforts}
%There are two types of efforts. 
%are of three types. 
are on three aspects. 
(1) Instrumenting a GPU driver to build a recorder.
%(1) To build a recorder for a GPU, developers instrument the GPU's driver. 
The effort is no more than 1K SLoC \textit{per GPU family}, as the instrumentation applies to the family of GPU SKUs supported by the driver. 
See \sect{recording} for examples. 
%The effort is often amortized over a family of GPU models supported by the GPU driver. 
(2) Recording their ML workloads. 
The effort is \textit{per GPU SKU}. 
%The effort can be further reduced: 
With minor patches, a recording can further be shared across GPU SKUs of the same family. 
% Section~\ref{sec:impl} describes our experiences. 
(3) Building a replayer. 
The effort is a few K SLoC \textit{per deployment environment}, e.g. for a TEE.

\subsection{The GPU Model}
\label{sec:abs_model}
%\jin{may need to add no in-flight power transition}

\sys{} builds on a small set of assumptions as summarized in Table~\ref{tab:gpus}.
As the ``least common denominator`` of modern integrated GPUs, 
the assumptions constrain GPU behaviors to be a reproducible subset. 
%they define modest knowledge of GPU interfaces as required by \sys{}. These are summarized in Table~\ref{tab:gpus}.
% Table~\ref{tab:gpus} summarizes how the model fits common integrated GPUs. 

\begin{myitemize}
\item 
% https://www.kernel.org/doc/html/latest/trace/mmiotrace.html
%\textbf{Registers and shared memory}
\textit{CPU/GPU interfaces} include memory-mapped registers, shared memory, and interrupts. 
Some GPUs, e.g. NVIDIA Tegra X1, may invoke DMA to access GPU registers~\cite{host1x}.
All these interactions can be captured at the driver level. 
% can capture at the CPU/GPU boundary. 

\item 
\textit{Synchronous job submission}.
% The CPU submits GPU jobs and learns their completions in lockstep. 
%The CPU submits a GPU job only after the prior job is completed.  % completes and learns their completions in lockstep. 
Disabling asynchronous jobs avoids interrupt coalescing and the resultant replay divergence.
% nondeterministic GPU execution. 
%The performance loss is modest as exemplified in Figure~\ref{fig:sync-overhead}.
The performance loss is modest as described in \sect{bkgnd:trend}.
%\sect{eval} will evaluate performance loss. 

\item 
\textit{GPU virtual memory}.
%GPU runs on its virtual addresses. % backed by its dedicated page tables. 
The replayer can manipulate the GPU page tables and load memory dumps to physical addresses of its choice.
% the replayer can relocate memory dumps at the replay time. 
% The recorder can dump GPU-visible memory and relocates the memory dumps at the replay time. 
%\sys{} \textit{can} work on legacy GPUs that run on physical memory. 
%Yet, the recorder must dump excessive memory and the replayer must run on the same physical memory range as the record time. 
\sys{} \textit{can} work with legacy GPUs running on physical memory.
Yet, the replayer must run on the same physical memory range as the record time. 
%the recorder must dump excessive memory and the replayer must run on the same physical memory range as the record time. 

\end{myitemize}

\begin{table}[]
\centering
\caption{Our GPU model fits popular integrated GPUs. *= To enforce sync job submission: Mali: reduce the job queue length; TegraX1: inject synchronization points to a command buffer; Adreno: check submitted job completion before a new command flush. NC: no changes}
\vspace{-5pt}
\fontsize{8.2}{8.7}\selectfont
\begin{tabular}{llll|llll}
                & \multicolumn{3}{l}{Features} & \multicolumn{4}{c}{Interface Knowledge}               \\ \hline
%                & MMIO    & Virt mem    & Sync job   & Jobstart & Pgtables & Reset & Irq Start/end \\ \hline
				 & \rotatebox[origin=c]{90}{MMIO}    & \rotatebox[origin=c]{90}{VirtMem}     & \rotatebox[origin=c]{90}{SyncJob*}   & \rotatebox[origin=c]{90}{JobStart} & \rotatebox[origin=c]{90}{Pgtables} & \rotatebox[origin=c]{90}{Reset} & \rotatebox[origin=c]{90}{IRQ} \\ \hline
Arm Mali~\cite{bifrost-driver}	& Y & Y & \cite{mali-sync}	& \cite{mali-submit} 	& \cite{mali-pgtable}	& \cite{mali-reset}      & \cite{mali-irq}              \\

%Mali Midgard~\cite{midgard-driver} 	& Y & Y & C  &          &          &       &               \\
 
%Bcom v3d~\cite{v3d-driver}        	& Y & Y & NC & \cite{v3d-jobstart-bcl,v3d-jobstart-rcl,v3d-jobstart-csd}        &     \cite{v3d-pgtable}     & \cite{v3d-reset}   &    \cite{v3d-irq}           \\
Bcom v3d~\cite{v3d-driver}        	& Y & Y & NC & \cite{v3d-jobstart-csd}        &     \cite{v3d-pgtable}     & \cite{v3d-reset}   &    \cite{v3d-irq}           \\

Bcom vc4~\cite{vc4-driver}        	& Y &   & NC & \cite{vc4-submitcl}	&    N/A      &  \cite{vc4-reset}     &       \cite{vc4-irq}        \\

%NV TegraX1~\cite{tegra-driver} 	& Y & Y & C	& \cite{tegra-submit}	& \cite{tegra-pgtable}	& \cite{tegra-do-idle,tegra-do-unidle}      & \cite{tegra-irq}              \\
NV TegraX1~\cite{tegra-driver} 		& Y & Y & \cite{tegra-sync}		& \cite{tegra-submit}	& \cite{tegra-pgtable}	& \cite{tegra-do-unidle}      & \cite{tegra-irq}              \\

Qcom Adreno~\cite{adreno-driver}	& Y & Y & \cite{adreno-sync}	& \cite{adreno-submit}	& \cite{adreno-pgtable}	& \cite{adreno-reset}	&  \cite{adreno-irq}             \\ \hline
\end{tabular}
%\vspace{3pt} % (\S\ref{sec:abs_model}) 

\label{tab:gpus}
%\vspace{-5pt}
\end{table}

%\subsection{Design considerations}
%\label{sec:design}

% !TeX root = main.tex

\begin{table*}[]
	\centering
	\caption{Replay actions in a recording}
	\vspace{-4pt}		% use as needed
	\includegraphics[width=0.98\textwidth{}]{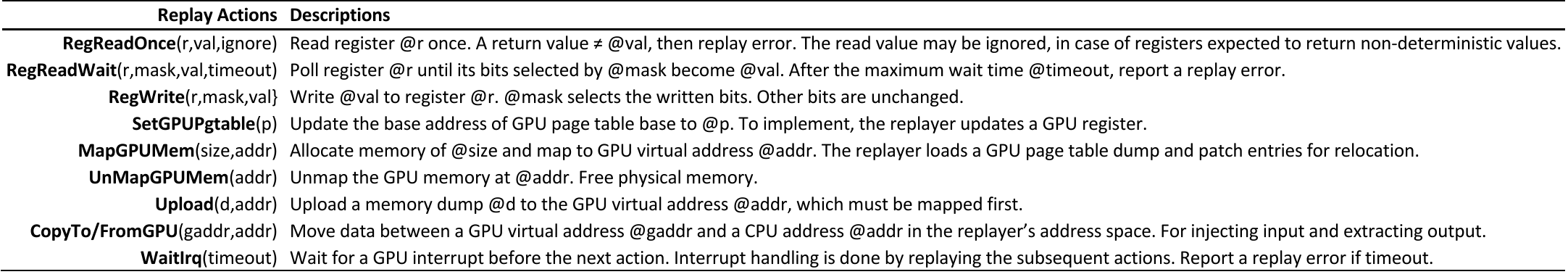} 
	%\missingfigure[figwidth=6cm]{Juno platform power instrumentation}
%	\vspace{-4pt}		% use as needed
	%\caption{The nondeterministic behavior of GPU and its software stack interaction}
	\label{tab:ir}
	\vspace{-4pt}		% use as needed
\end{table*}

%\subsection{Replay correctness}
\paragraph{Replay correctness}
%\noindent
%\textbf{Replay correctness}
%The replayer offers the same guarantee for detecting incorrect execution as the full driver does. 
The replayer offers the same level of correctness guarantee as the full GPU stack does: 
%the replayer's assertion on successful completion of a ML workload is as sound as an assertion from the GPU stack. 
the replayer's assertion that a recorded workload (a series of GPU jobs) is completed is as sound as an assertion from the GPU stack. 
%Our rationale is as follows. 
Our rationale is based on the GPU state. 
%We have the following observation. 

A GPU state ${<P,C,J>}$ is all GPU-visible information affecting the GPU's execution outcome: P is the GPU's current protocol step, e.g. wait for commands; C is the GPU's hardware configuration; J is the job binary being executed. 
\textit{We define a replay run as correct} if the GPU at the replay time goes through the same state transitions as the record time. 
%Therefore, if the GPU at the replay time goes through the same state transitions as recorded, the GPU must have completed the same workload (albeit on different input). \textit{We define such a replay run as correct}. 

%all the GPU-visible memory state. 
%A GPU workload is uniquely characterized by the state transitions the GPU goes through during executing the workload. 
%A GPU workload is uniquely characterized by the GPU state transitions during the workload execution. 
%and the job binaries the GPU currently runs on.  
%If in a replay the GPU goes through the same state transitions as the record time, we consider the replay is correct. 

%\noindent
%\textit{State-changing events.}
%A GPU driver continuously tracks GPU state to assess if the current execution is correct. 
The full GPU driver, as it runs, continuously assesses if the GPU state deviates from a correct transition path. 
%It does so solely based on observation of \textit{state-changing} events in CPU/GPU interactions. 
The driver's only observations are \textit{state-changing} events in CPU/GPU interactions: the events either changing the GPU state or indicating the GPU state has changed. 
State-changing events include: 
a register write; 
a register read returning a value different from the most recent read; 
a register read with side effect; 
interrupts. 

%The replayer reproduces CPU/GPU interactions. 
Based on the rationale, the replayer asserts correctness based on matching state-changing events. 
If it observes the same sequence of state-changing events with all event parameters matched, 
%then as far as the full GPU driver is concerned, 
then to the best knowledge of the GPU driver, 
the GPU makes the same state transitions and completes the recorded workload.  
The replay is correct per our definition. 

Suppose a state divergence, such as silent data corruption, is missed by the replayer, it could have been missed by the full GPU driver as well. 
If we assume the driver is \textit{gold}, i.e. it has made sufficient interactions to assess if GPU state has deviated from the correct transitions, 
% if the GPU state has deviated from 
%is deviating from the correct transitions, 
then such silent divergences should neither occur to the driver nor the replayer. 

\paragraph{Nondeterministic CPU/GPU interaction}
Even to repeat the same workload, the CPU/GPU interactions are likely to differ, e.g. 
CPU may observe diverging register values or receive extra/few interrupts. 
%CPU may observe diverging register values or receive extra/few interrupts. 
%and sees unexpected GPU memory exceptions. 
%The replayer, hence, cannot use raw traces.
Hence, a raw trace cannot be replayed verbatim.
%This means the replayer cannot replay raw traces verbatim. 
The major nondeterminism sources are as follows.
%We identify the following nondeterminism sources 
(1) Timing. 
For instance, a GPU job's delay may vary; 
the CPU may poll the same register for different times until its value changes. 
% and in acknowledging GPU interrupts may vary across runs. 
%Examples include polling a register for different times. 
(2) GPU concurrency. 
The order of finishing concurrent jobs and the number of completion interrupts may vary. % , e.g. due to coalescing. 
(3) Chip-level hardware resources, e.g. changes in a GPU's clockrate. 
% , which may be subject to SoC-level thermal throttling. 

Because replay correctness only depends on GPU states, 
we treat nondeterminism as follows. 
%
%(1) Nondeterministic interactions that do not diverge GPU state: 
%(1) Nondeterminism not affecting GPU state. 
(1) Nondeterminism not affecting GPU states. 
This includes most of the timing-related behaviors. % nondeterminism. 
The recorder discovers and summarizes them as replay actions, 
so that the replayer can tolerate (\S\ref{sec:recording}).  
% avoids imposing unnecessary constraints on interaction events. 
%(2) Nondeterminism affecting GPU state but preventable.
(2) Affecting GPU states; preventable.
This includes GPU concurrency and some configurable chip resources. 
We eliminate the nondeterminism sources, e.g. enforcing synchronous job submission as described in the GPU model above. 
% and disallowing runtime power management. 
%turning off aggressive power management; 
%We eliminate the sources of such nondeterminism by constraining GPU dynamism (see section~\ref{sec:abs_model})
%turning off aggressive power management; 
%(3) Nondeterminism affecting GPU state and non-preventable. 
(3) Affecting GPU states; non-preventable.
This mainly includes strong contention and failures in chip resources,
such as power failures. % an undercurrent condition. 
The replayer detects them and attempts re-execution.

\begin{comment}

GPU dynamism leads to replay divergence: different register values or interrupts, unexpected GPU memory exceptions, and most commonly, no response. 
The replayer, lacking intimate GPU knowledge, is unable to resolve the divergence. 
GPU dynamism must be constrained at the record time. 

The causes include 
(1) Non-deterministic job delays, subject to memory/interconnect contention and GPU's shader core scheduling; 
(2) IRQ coalescing. A GPU may opportunistically batch job completion notifications~\cite{genesys, basu18iiswc}. % to reduce interrupts.
% We observed SqueezeNet example in ACL submits the 71 jobs but receives 45 interrupts.
% and expect the CPU to learn what were completed from register reads~\cite{genesys, basu18iiswc}. 
% The number of interrupts received by CPU depends on GPU job timing. 
(3) Power management. GPU may sleep after a period of inactivity on its own. 
% Waking up the CPU will incur delays or require additional register accesses. 
CPU timing, e.g. varying delays in acknowledging GPU interrupts, further adds to the dynamism. 
% in submitting jobs to GPU after they are prepared, or in polling of GPU registers. 
\end{comment}

% !TeX root = main.tex

\section{Record}
\label{sec:recording}

% \input{fig-mmap}
% \input{fig-symbol-inout}
%What to record? - the information must be enough to reflect all the necessary things forming IR.
%blackbox user-space runtime challenges: unknown in/output address
%online (on-demand) behavior of the GPU: from granularity - dynamic memory allocation, page table update (pte)

%To address them, we illustrate a recording guide by symbol-based address identification and dynamic memory snapshot.
%We list necessary information to capture for reproducing computation and describe where to capture. (\S\ref{sec:recording_trace}).
%We unveils unknown in/output address space by the symbol-based address space identifica￩tion (\S\ref{sec:recording_trace}).
%We dynamically captures address space (\S\ref{sec:recording_trace}).

%We build the recorder by instrumenting a GPU driver at the following code locations: 

\subsection{Interface Knowledge and Instrumentation}
%\paragraph{Known interface semantics}
%\paragraph{Interface knowledge}
% We assume the following GPU interface semantics are known. 
% \sys{} relies on the knowledge of a subset of GPU interfaces. 
%We rely on the following GPU interface knowledge as summarized in Table~\ref{tab:gpus}. 

The knowledge needed by the recorder is in Table~\ref{tab:gpus}: 

\begin{myitemize}
\item 
The registers for starting a GPU job and for resetting GPU.
%The register for starting a GPU job, which allows the recorder to trigger GPU memory dumping.
%The register for resetting GPU, which allows clean-slate GPU handoff before and after replay.  

%\item 
%The registers for starting a GPU job and for resetting the GPU. 
%This allows the recorder to trigger GPU memory dumping
%The register . This allows clean-slate GPU handoff before and after replay.  

\item The register pointing to the GPU page tables; 
the GPU page table's encoding for physical addresses. 
%the bits in a GPU page table entry encoding the physical address. 
%This allows the recorder to capture the GPU virtual address space and the replayer to set up the address space. 
This allows to capture and restore the GPU virtual address space. 

% \item The register for resetting the GPU. This allows clean-slate GPU handoff before and after replay.  

\item The set of registers on which reads or writes do not change GPU state. 
This is to detect state-changing events. 

\item The events that a GPU interrupt handler starts and ends.  
Knowing them allows the replayer to enter and leave an interrupt context (via \texttt{eret}) just as the record time. 

\item (Optional) The events that the GPU hardware becomes busy or idle. 
The recorder uses them to remove unwanted delays. 
%Knowing them allows the recorder to remove unwanted delays from recording. 
\end{myitemize}

%\paragraph{How much developer effort?}
%\paragraph{Developer efforts}
%Table~\ref{tab:gpus} summarizes the interface knowledge for popular GPUs, which we extract from their open-source drivers. 

%The efforts to extract the above knowledge, as well the efforts to instrument a driver, are often shared by a family of GPU models supported by the driver. 
%We confirm this is true for 6 GPU models supported by the Arm Bifrost driver~\cite{bifrost-driver} and 17 GPU models supported by the Adreno 6xx driver~\cite{adreno-driver}. 
%% Although a driver may execute code conditionally depending on the GPU model in use, the register maps and semantics of a GPU family are almost identical. 
%Although a driver may execute code conditionally depending on the GPU model in use, most register names and semantics of a GPU family (>XX\% for Bifrost and >XX\% for Adreno) are identical. 

%We instrument the following driver code locations: 
We instrument the driver code: 
register accessors; % in order to capture register accesses; 
register writes starting a GPU job;  % in order to trigger memory dumps; 
accessors of GPU page tables; % in order to capture page table updates; 
interrupt handling.  %in order to capture interrupt start/end. 
Many of these code locations are already abstracted as macros~\cite{v3d-regs} or tracepoints~\cite{mali-tracepoint}. 
%Existing drivers already abstract register accessors as macros~\cite{v3d-regs} and define tracepoints~\cite{mali-tracepoint} for most of these locations. 
% As a result, we only have to change as few as hundreds lines of driver code (e.g. 500 SLoC for Mali Bifrost).
%We only add a few hundred lines of code per driver. 
% We find manual instrumentation is more robust than a transparent approach, e.g. tracing register accesses via page faults~\cite{mmio-trace}.
We find manual instrumentation is more robust than tracing via page faults~\cite{mmio-trace}.

%Once activated, the recorder enforces synchronous job submission. 
%This requires either no changes or light changes to the GPU driver, as summarized in Table~\ref{tab:gpus}.

% This is already the default behavior of some GPU drivers (e.g. v3d) so no changes are needed. 
% For other drivers that allow multiple outstanding jobs, we add a few lines of code that sets the job queue length to be one.  Changing the driver is cleaner than app-level methods, e.g. injecting CLFlush() calls from the record harness.

\paragraph{Developer efforts}
to extract interface knowledge and to instrument a driver are often amortized over a family of GPU SKUs supported by the driver. 
We confirm this is true for 6 GPU SKUs supported by the Arm Bifrost driver~\cite{bifrost-driver} and 17 GPU SKUs supported by the Adreno 6xx driver~\cite{adreno-driver}. 
% Although a driver may execute code conditionally depending on the GPU model in use, the register maps and semantics of a GPU family are almost identical. 
Although a driver may execute code conditionally depending on the GPU SKUs in use, the GPU interfaces in a GPU family, 
i.e. register names and semantics, are often identical.

%\textit{i) the power is always on}: we advocate a static power policy that always turns the GPU on during recording. Note that the power is configurable as it is managed by device driver level in general. Thereby, no power transition (suspend/resume) on the GPU compute cores occurs during recording.

%\subsection{The IR of GPU recording}
%\subsection{The recording format (for XXX..)}
\subsection{Register Access}

A recording consists of actions listed in Table~\ref{tab:ir}. 
%An action may summarize a sequence of register accesses in order to tolerate nondeterminism within. 
An action may summarize a sequence of register accesses showing nondeterminism without affecting GPU state. 
For instance, CPU may wait for GPU cache flush by polling a register ~\cite{v3d-cleancache, mali-cleancache}, where 
the number of register reads depends on the nondeterministic flush delay. 
%the replayer will likely read different times than what was recorded.
Such polling is summarized by RegReadWait(). 
% An action RegReadWait() hides such nondeterminism. 

To do the above,
the recorder recognizes nondeterministic register accesses that do not change GPU state. 
With the GPU interface knowledge described above,
we inspect a driver's register accessors and instrument their callsites that match the patterns in Table~\ref{tab:ir}. 
%The effort is tractable: multiple drivers already abstract these accesses with macros, e.g. wait\_for()~\cite{v3d-waitfor,tegra-wait-for-idle}.
We tap in existing macros such as wait\_for()~\cite{v3d-waitfor,tegra-wait-for-idle}
and instrument tens of callsites per driver. 
% The effort is tractable: there are only several common patterns (see Table~\ref{tab:ir});  we only have to inspect and instrument XXX callsites per driver (XXX for XXX, XXX for XXX);  some drivers already use macros to abstract these register access patterns, e.g. wait\_for() ... 

\begin{comment}
A recoding refers to GPU registers by names instead of by addresses. 
For legal register access, the target app trusts the replayer rather than trusting a recording's integrity. 
% for containing legitimate addresses. 
%This allows the target app to put trust on its replayer rather than trusting recordings for containing legitimate addresses. 
Section~\ref{sec:eval-security} will discuss the security implications.
\end{comment}

\subsection{Dumping Proprietary Job Binaries}
The recorder must record for a job's binary: 
(1) GPU commands for data copy or format conversion, often packed as nested arrays;  
(2) shaders, which include GPU code and metadata; % e.g. shape of thread groups; 
(3) GPU page tables. % which form the virtual address space for the job. 
% At run time, the GPU runtime sets up the GPU address space and accordingly links commands and shaders to GPU addresses. 
%
A GPU binary is deeply linked against GPU virtual addresses: 
GPU commands contain pointers to each other, to the shader code, and to a job's input data; 
shaders also reference to code and data.
%shaders also contain pointers to code and data. 
% the shaders also reference to code and data at their GPU addresses. 
% Once a GPU binary is emitted, these GPU virtual addresses cannot be changed without intimate GPU knowledge. % in general. 
% \sys{} has no knowledge to update GPU virtual addresses in an emitted GPU binary. 
Therefore, \sys{} dumps all memory regions that may contain the job binary; to replay, \sys{} restores the memory regions at their respective GPU virtual addresses. 
% It does not touch GPU virtual addresses in an emitted GPU binary. 

% A recoding contains only GPU virtual addresses but not physical addresses. This makes the replayer solely responsible for legitimate memory accesses regardless of recording integrity. 

\paragraph{Time the dump}
% To record a workload, the recorder dumps job binaries in multiple shots. 
%A GPU stack emits job binaries lazily and updates GPU page tables for upcoming jobs incrementally. 
A GPU stack emits a job's binaries and updates GPU page tables lazily -- often not until it is about to submit the job. 
%Accordingly, the recorder dumps GPU memory right before the driver writes to the register that starts each new job.
Accordingly, the recorder dumps GPU memory right before the driver kicks the GPU for a new job. 
At this moment, the runtime must have emitted the job's binary to the GPU memory; 
% which is ready to dump. 
the memory dump must be consistent: synchronous job submission ensures no other GPU jobs are running at this time and mutating the memory.

\paragraph{Locating job binaries in GPU memory}
Memory dumps must include job binaries for correctness;
% For instance, they should exclude GPU's intermediate buffers, which are sizable and will be overwritten at replay time anyway.
% They should exclude a job's input buffer (which may be output of a previous job) 
they should exclude GPU buffers passed among jobs so that loading of memory dumps does not overwrite these buffers; 
they should leave out a job's scratch buffers as many as possible for space efficiency. 
% , which are sizable. 
% which are sizable and will be overwritten at replay time anyway.

The challenge is that the recorder does not know exactly where GPU binaries are in memory: 
%where GPU commands and shaders are in memory. 
the GPU runtime directly emits the binaries to mmap'd GPU memory, bypassing the GPU driver and our recorder therein. 
A naive dump capturing all physical memory assigned to GPU can be as large as GBs. 
An optimization is to only dump memory mapped to GPU at the moment of job submission, which reduces a memory dump to MBs. 
%To do so, the recorder parses GPU page tables and identifies such GPU regions. 
% The recorder identifies mapped GPU regions by (1) parsing the current GPU page tables, or (2) instrumenting the driver function that updates GPU page entries. 
% Both methods fit in our GPU model. 
% We implemented them on concrete GPUs with low engineering effort, as will be shown in Sec XXX. 
%A dump only containing GPU-visible memory may still contain unwanted data. 
\sect{impl} presents hardware-specific optimizations to further shrink memory dumps. % with heuristics on GPU registers and page tables.  

\subsection{Locating Input and Output for a Recording}
\label{subsec:locating_inout}

\paragraph{Record by value vs. by address}
A recording accepts one or more input buffers. 
By default, \sys{} records an input buffer by address: 
the recorder captures the buffer's GPU address, allowing new data injected at the address at replay time. 
Use cases include an NN's input buffer. 
If developers intend to reuse an input buffer's values for replay, they
may optionally annotate the input as ``record by value'' in the record harness. 
\sys{} then captures the buffer values as part of memory dumps. 
Use cases include a buffer of NN parameters. 
An input recorded by value and by address simultaneously allows \textit{optional} value overriding. % at replay time. 
Annotations only decide apps' responsibility for providing input data at the replay time; 
improper annotations do not break replay correctness.
%Incorrect annotations does not affect replay correctness; yet, ML apps have to provide new values for certain inputs and/or can only use old values as recorded for certain inputs. 
% If developers make incorrect annotations, their ML app either have to provide new values for certain inputs additionally or can only use old values for certain inputs. 
% lose the ability to override the values of certain inputs. 
% The replay correctness is not affected. 
%The recorder always captures a recording's output by address, from where the replayer will extract the output data.
%  e.g. an NN's output class label. 

% Note that \sys{} remains oblivious to inputs/output of individual jobs in a recording, as a recorded GPU job binary already encodes the knowledge on locating its input from the previously job's output. 

\begin{comment}
\Note{TODO: reword. perhaps a figure in S3 can help}
Note that \sys{} is oblivious to, and does not capture, the inputs/output for jobs \textit{internal} to a recording. 
This is because the recorded GPU jobs already know how to locate their input/output in the GPU address space. 
\end{comment}

\paragraph{Discover input/output addresses}
Recording \textit{by value} is straightforward: 
just dump any memory region that \textit{may} contain the input.
% Loaded prior to executing all the jobs, the dump will not overwrite intermediate buffers shared among jobs. 
% Since the dump will be loaded prior to executing all the jobs, 
% including unwanted data in the dump will not affect replay correctness.
Recording \textit{by address} is more challenging: 
the recorder cannot track to which GPU address the runtime copies input, as the runtime is a kernel-bypassing blackbox;
it does not know from which addresses the GPU code loads input, because the recorder cannot interpret the GPU code. 
% as the recorder cannot parse the GPU code that encodes or calculates these addresses. 

To reveal these memory locations, \sys{} adopts a simple taint tracking. 
The record harness injects input magic values -- synthetic, high-entropy data -- and looks for them in GPU memory dumps. 
The rationale is that it is very unlikely that a high-entropy input (e.g. a 64x64 matrix with random elements) coincides another GPU memory region with identical values.

%The rationale is that it is very unlikely that a high-entropy input (e.g. a 64x64 matrix filled with random values) coincides another GPU memory region with identical values.

% Before the first job and after the last job,  the recorder dumps all mapped GPU memory (not only the job binaries) which must include the input and output magics. 
% From the offsets of magic values found in the memory dumps, the recorder calculates the GPU addresses of input/output. 
%the recorder takes into the then GPU address spaces into account and  calculates the GPU addresses of input/output. 

We took care of a few caveats. 
(1) The output often has lower entropy because it is smaller (e.g. a class label). 
% jin:conflict
%(1) The output often is smaller and has lower entropy (e.g. a class label). 
In case of multiple matches of output magic in memory, 
\sys{} repeats runs with different input magics to eliminate false matches. 
%(2) \sys{} may fail to locate the magic values if CPU transforms an input before sending it to the GPU memory, e.g. to reshape the input data. 
(2) The above technique cannot handle the case when the ML framework runs CPU code to reshape data before/after the data is moved to/from GPU. 
%sitting between apps and GPU runtime, reshape input/output using CPU code. 
Fortunately, we did not see such a behavior in popular ML frameworks: Tensorflow, ncnn, and ACL. 
%For efficiency, they always keep data in the GPU region and invoke GPU, if available, for data reshaping.
For efficiency, they always invoke GPU, if available, for data reshaping.
% Therefore, our recorder can always locate the input in the GPU memory. 
% In case CPU indeed transforms the input, our recorder would have to track magic values at finer granularities with sophisticated taint tracking \Note{cite}. 
While we are aware of rigorous, fine-grained taint tracking~\cite{tupni}, our simpler technique is sufficient for locating GPU input/output. 
This saves us from configuring symbolic execution on a closed-source GPU runtime of tens of MBs, which requires expertise and non-trivial effort.

\subsection{Pace Replay Actions}
\label{subsec:delay_reduction}

% hence pacing the replay.

CPU cannot replay as fast as possible, otherwise GPU may fail to catch up.
For example, CPU needs to delay after resetting the GPU clock/power for them to stabilize~\cite{tegra-unrailgate,mali-pm-policy} and delay after requesting  GPU to flush cache~\cite{tegra-l2flush}.

% CPU needs to wait for tens of us for a GPU cache flush to complete. 

% As an example, CPU needs to wait for 1 us for changes to GPU clocks to propagate \Note{cite}; 
% CPU needs to wait for tens of us for a GPU cache flush to complete. 
% For instance, it takes 60 us, an empirical delay, for GPU cache flush in Mali Bifrost G71.
%For instance, \Note{60 us, an empirical delay for GPU TLB flush? (v3d)}. 

%To pace the replay, 
The recorder sets a minimum interval $T$ for each action: 
if the replayer takes $t$ to execute the current action, it pauses for at least $T-t$ to before the next action. 
Setting proper intervals is non-trivial. 
% At the record time, 
When running the GPU stack, 
CPU paces its interactions with GPU intentionally (e.g. calling delay()) or unintentionally (e.g. running unrelated apps). 
% the full GPU stack paces CPU/GPU interactions, intentionally (e.g. the driver calling delay()) or unintentionally (e.g. running other apps). 
% The resultant delays give the GPU enough time to respond, 
The recorder should not preserve the observed intervals, as 
doing so will unnecessarily slow down the replay. 
%resulting in performance loss. 

% !TeX root = main.tex

\begin{figure}
	\centering
	\vspace{1pt}
	\includegraphics[width=0.47\textwidth{}]{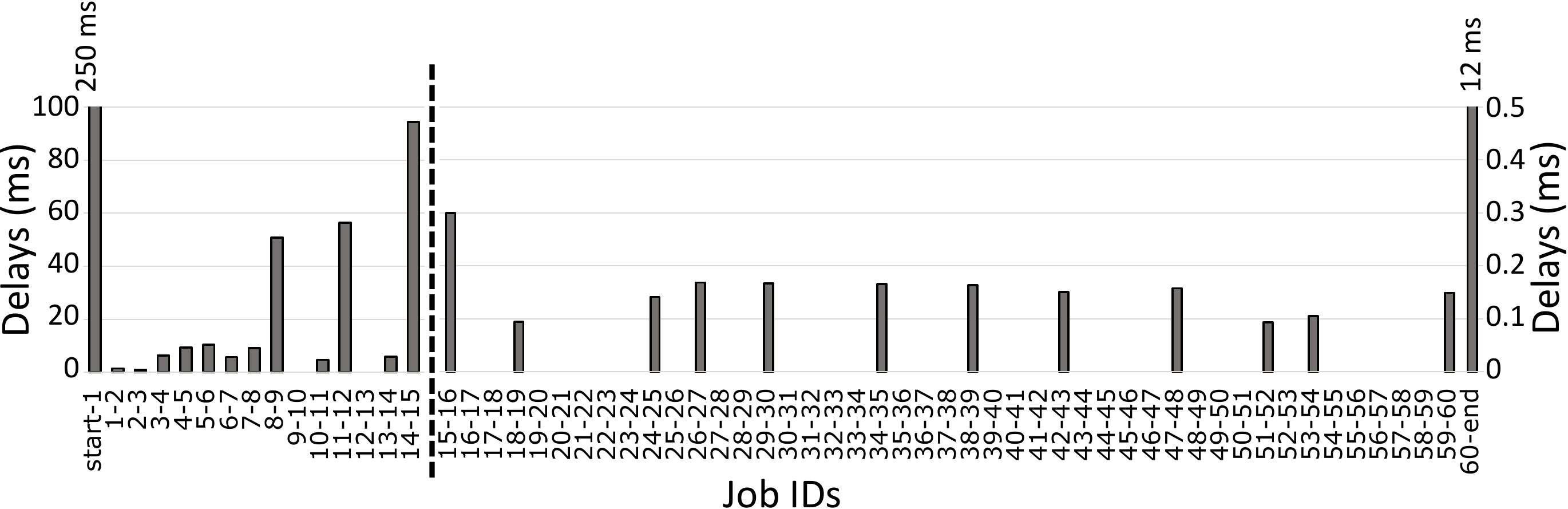} 
	%\missingfigure[figwidth=6cm]{Juno platform power instrumentation}
%	\vspace{-5pt}		% use as needed
%	\caption{User delay distribution of AlexNet. 1 -- 30 ms unnecessary delays are added before each compute kernel execution.}
%	\caption{Intervals between CPU/GPU interactions, accumulated on individual GPU jobs. Workload: AlexNet inference. Mali G71 on ACL}
	\caption{Intervals between CPU/GPU interactions, accumulated by GPU job. Intervals among earlier jobs are longer than later ones. Workload: AlexNet inference. ACL~\cite{acl} on Mali G71. Excluded: GPU busy time; parameters loading IO}
	\label{fig:nondeterminism}
	\vspace{-4pt}		% use as needed
\end{figure}

Figure~\ref{fig:nondeterminism} shows an example, 
where most long intervals are unintended delays from CPU: 
% Most long intervals occur near the execution's start and end. 
(1) 
Resource management, such as initialization of GPU memory management; 
% including frequent CPU/GPU memory allocation and memory mapping. 
(2) JIT generation of GPU commands and shaders;
(3) OS asynchrony, 
such as scheduling delays; 
% such as delays in a user-level thread pool and in scheduling kernel threads for GPU job submission;  
%such as delays in scheduling a thread pool in an ML framework, 
%and in scheduling kernel threads for job submission and completion. 
%An ML framework with a thread pool \Note{cite} can delay submitting GPU jobs when if thread workers are busy. \Note{example?}
%The kernel thread for submitting GPU jobs may be run tens of milliseconds \textit{after} a job is ready. 	
% For efficiency, the stack often delays job submission and notifications by deferring them to worker threads. 
(4) Recording overhead, e.g. dumping GPU memory; 
% e.g. creating a memory dump often takes tens of milliseconds. 
(5) Abstraction tax, e.g. frequent IOCTLs. % syscalls. 
% (5) Abstraction tax for supporting a variety of programming abstractions and GPU hardware, e.g. frequent user/kernel crossings. 
% We measured that a XXX simple Vulkan API will call XXX levels of functions before reaching the user/kernel boundary. 
% User/kernel boundary crossings (e.g. XXX IOCTL syscalls) further add to the overhead. 
Doing none of these, 
the replayer should simply skip the resultant intervals and 
fast-forward to the next action. 
% The delays accumulate to XXX\% of the total recording length. 
%The challenge is to identify unintentional slowdown. 
% The challenge is to differentiate unintentional slowdown from observed intervals. 

The challenge is to differentiate unintended delays from intended delays.
% Even on a simple workload, the GPU stack runs multiple threads with complex interleaving. 
% At the record time, 
% As the intervals observed by the recorder are from interleaving of multiple CPU threads, it is unrealistic for the recorder to profile the GPU stack and find out interval causes. 
% It is unrealistic for the recorder to profile the GPU stack running multiple threads to exercise the GPU in order to reveal the interval causes. 
It is unrealistic for the recorder to profile the complex, multi-threaded GPU stack. % and reveal the causes of delays. 
Instead, it follows a simple heuristics: 
\textit{if the GPU hardware has been idle through one interval, the interval is safely skippable}. 
The rationale is that an idle GPU can always keep up with CPU's next action without pause. 
% For example, after the GPU completes a job J0, the CPU takes another 10 ms to compile and submit the next job J1. 
% Since the recorder can prove that GPU is idle in the 10 ms interval, it safely excludes the interval from recording. 
%
%it safely excludes the interval from recording. 
%For example, after a GPU interrupt signaling job completion, the CPU takes extra 100 ms to finish compiling and then submits the next job. 
%Since the recorder can prove that GPU has been idle in the 100 ms interval, 
%it safely excludes the interval from recording. 
%
With this heuristics, we add tens of lines of code per driver, which can prove GPU idle for more than half of the observed intervals. Skipping them speeds up the replay significantly, as we will show in \sect{eval}. 
The recorder simply preserves the remaining intervals for replay.
\section{Replay}
\label{sec:replayer}

The replayer provides the following APIs. 
(1) \textit{Init/Cleanup}: acquire or release the GPU with reset.
(2) \textit{Load}: load a recording file, verify its security properties, and allocate the required GPU memory. 
(3) \textit{Replay}: replay the recording with input/output buffers supplied by the app. 
%If it succeeds, the output buffer will contain the results. 
The replayer consists of a static verifier; 
an interpreter that parses/executes a recording in sequence; 
a nano GPU driver to be invoked by the interpreter. 
% We next present their designs. 
% It maps GPU registers, allocates the needed amount of physical memory, and ... For a record entry updating the GPU memory mapping, the replayer invokes the nano driver to update the corresponding page table entries. 

\begin{comment}
\item Init/Cleanup. Acquire or release the GPU with reset.
\item Load. Load a recording file, verify its security properties, and allocate the required GPU memory. 
\item Replay. Replay a loaded recording with input and output buffers supplied by the app. 
If it succeeds, the output buffer will contain the compute results. 
\end{comment}

\subsection{Verification of Security Properties}
\label{sec:verficiation}
% The app and the replayer do not have to trust a recording. 
% A target app trusts the replayer which statically verifies the following security properties and enforces the properties at run time. 

%The target app does not have to rely on a recording's integrity. 
The replayer statically verifies the following security properties. 
While a full GPU driver may implement similar checks, the replayer provides stronger guarantees due to its simplicity and independence of an OS kernel. 
%For these security properties, 
%The target app hence trusts the replayer for these security properties rather 

\begin{myitemize}
\item 
%\textbf{Recording only accesses allowable GPU registers} 
\textit{No illegal GPU register access by CPU.} 
A recording contains GPU register names, which are 
resolved by the replayer as addresses based on the CPU memory mapping. 
%The replayer ensures that all registers referenced in a recording are legal GPU registers. 
%The replayer resolves the register names as addresses based on the CPU memory mapping. 

\item 
%\textbf{GPU only accesses its own memory}
\textit{No illegal memory access by GPU.}
%A recording never specifies physical addresses accessible to GPU. 
%Instead, it only specifies the sizes and GPU virtual addresses of these regions. 
A recording only specifies sizes and GPU addresses of memory regions. 
It is up to the replayer to allocate the underlying physical pages and set up GPU page tables. 
The replayer ensures the allocated physical pages contain no sensitive data. 
The GPU MMU prevents GPU code from accessing any CPU memory. 

\item 
%\textbf{The maximum amount of physical memory needed.}
\textit{Maximum GPU physical memory usage.}
%By its hardware design, a GPU can only access memory mapped to it. 
%All GPU-visible memory must be mapped before use. 
The replayer scans a recording for \texttt{MapGpuMem} entries (Table~\ref{tab:ir}) to determine the GPU memory usage at any given moment. 
Based on the result, apps or the replayer can reject memory-hungry recordings. % that may render the whole system in low memory. 
%that threatens the whole system's memory availability. 
%With the knowledge, the replayer can ensure a replay never fails in the middle due to out of memory. 
%It can reject 

\end{myitemize}

%The replayer cannot decide if a recording is semantically correct, which is an orthogonal concern.
The replayer cannot decide semantic correctness which is orthogonal to security.
\sect{eval-security} will present discussions.

\begin{comment}
The replayer, however, cannot statically bound a recording's execution time, as it lacks the knowledge of GPU shaders and commands. 
The replayer detects GPU hang with a CPU-side watchdog timer, and handles hangs as replay failures~(\S\ref{sec:failure-handle}).
\end{comment}

%When a user in the target device requests, the \rep{} first verifies whether \ir{} in the package is compatible with its own GPU (e.g. read the GPU ID).
%\jin{check/verification -- verifying register are GPU regs, memory regions are legal, etc}
%When a user in the target device requests, the \rep{} first verifies whether \ir{} in the package is compatible with the its own GPU (e.g. read and compare GPU ID). Note that no GPU preemption is allowed during one replaying (\S\ref{sec:abs-model}). Then, the \rep{} remaps base address register of GPU (or MMIO region) to its own address space (e.g. kernel memory) like as what \texttt{ioremap} does and cleans up all the previous GPU states.

%Before replaying, the \rep{} should check if the IOIR is compatible to its own GPU. To this end, the \rec{} passes the register read IOIR that accesses the register returning \texttt{GPU\_ID} together with the \texttt{GPU\_ID (rec)} used in recording. Then the \rep{} gets its own GPU ID, \texttt{GPU\_ID (rep)}, by register access. By comparing \texttt{GPU\_ID (rep)} to \texttt{GPU\_ID (rec)}, the \rep{} can verifies the record is compatible to its own GPU.

\subsection{The Nano GPU Driver}
The nano driver abstracts GPU hardware;
it only has of 600 SLoC. 
Most driver functions directly map to replay actions: mapping GPU registers to CPU addresses, 
copying data in and out of GPU memory, 
rewriting the GPU page table entries for loading memory dumps, etc. 
%In our implementation, it consists of only a few hundred lines of SLoC. 
%  much smaller than a full GPU driver which is up to tens of thousands of SLoC. 
% so that they point to the physical memory allocated at the replay time. 
%\paragraph{Interrupt handling}
The driver includes a bare minimum interrupt handler, 
% The interrupt handler itself 
% of interrupt handling: 
which simply switches the CPU to the interrupt context and continues to replay the subsequent actions. 
% when reaching an ``interrupt return'' action, the handler switches the CPU to the normal context. 
% it is invoked in response to a GPU interrupt, and then continues to replay the subsequent records; when it reaches an ``interrupt return'' record, the interrupt handler returns 
The interrupt management, such as waiting for an interrupt, acknowledging an interrupt, and checking interrupt sources, is done implicitly by replaying the corresponding actions.
% To implement a user-level replayer which cannot register interrupt handlers, the replayer polls for interrupt status through file descriptors, as will be described in Section~\ref{sec:replayer-env}. 
 
%\begin{myenumerate}%	
%	\item Timing: each IOIR contains timestamp which allows to maintain the same timing of register IO when replaying.
%	
%	\item Reg polling: let \rep{} stay in busy loop while polling the register (e.g. GPU and MMU cache flush \jin{(nondeterministic?)})
%\end{myenumerate}

\subsection{GPU Handoff and Preemption}
\label{sec:replay:handoff}
During replay, the replayer fully owns the GPU and does not share with other apps.
Before and after a replay, it soft-resets the GPU, ensuring the GPU starts from a clean state without data leaking, e.g. no subsequent apps will see unflushed GPU cache.
The replayer allows the OS to reset and preempt the GPU at any time (e.g. yielding to an interactive app) without waiting for ongoing GPU jobs to complete.
%The replayer allows the OS to preempt GPU at any time, e.g. yielding to an interactive app.
%The OS resets the GPU without waiting for any ongoing GPU job to complete.
Hence, preemption incurs short delays.
%Hence, preemption incurs short delays as reported in Section~\ref{sec:eval-design}. 
A preemption disrupts the current replay. 
To mitigate it, 
%To mitigate disruption, 
we implement optional checkpointing: periodically making copies of GPU memory and registers.
A disrupted replay resume from the most recent checkpoint.
%A disrupted replay can resume from the most recent checkpoint.
%The checkpoint interval is user configurable. 
% Compared to re-executing from the recording start, checkpointing benefits some, but not all, workloads. 
% Section~\ref{sec:eval-design} will evaluate the overhead and the impact of checkpointing intervals.  
\sect{eval} evaluates preemption and checkpointing experimentally.
%\sect{eval-design} will evaluate preemption delays and checkpointing experimentally. 
% \sect{eval} evaluates checkpointing and compares its overhead to re-executing from the recording start. 

%memcpy overhead may not be prnounced. 
%the problem canbe mem space. 
%but has to flush GPU cache every time. that could be bad. 
%GPU registers?
%Page table? 

% \input{fig-pgt}

%\subsection{Failure handling}
%\subsection{Handling state divergence}
\subsection{Handling Replay Failures}
\label{sec:failure-handle}

Replay failures are GPU state divergences due to non-preventable nondeterminism at run time. 
Based on our GPU model~(\S\ref{sec:overview}),
the replayer will not miss detecting any state divergences the full GPU stack can detect.
When the replayer faces failures, it attempts to recover through re-execution:
resetting the GPU and starting over the whole recording;
if the divergence persists, the replayer injects additional delay to the action intervals that precede the divergence occurrence.

\begin{comment}
Replay failures are GPU state divergences due to non-preventable nondeterminism at run time. 
Based on our GPU model in~\sect{overview}, 
%the replayer is as capable as the full GPU stack at detecting state divergence. 
%the replayer can detect all state divergences that the full GPU stack can detect. 
the replayer will not miss detecting any state divergences the full GPU stack can detect. 

The replayer attempts to recover through re-execution:
resetting the GPU and starting over the whole recording;
if the divergence persists, the replayer injects additional delay to the action intervals that precede the divergence occurrence.
\end{comment}

Re-execution with delays can overcome transient failures and many timing-related failures, which are the most common failures based on the driver code comments, documentations, and our own experience. 
Examples include 
an underclocked GPU for replay fails to keep up with the replay actions; 
high contention on shared memory cause GPU jobs to timeout. 
%(2) all the CPUs are under tremendous pressure, which hinders the replayer from timely reproducing state-changing events;

Re-execution cannot overcome persistent failures, e.g. reoccurring hardware errors. 
A full driver is unlikely to overcome such errors either. 
% Such errors, however, are unlikely to be overcome by a full driver either. 
In this case, the replayer seeks to emit meaningful errors as the full driver does: it reports the failed action and the associated source locations in the full driver.

\begin{table}[]
	\fontsize{9.0}{12.0}\selectfont
	\begin{center}
		
	\caption{\sys{} implementations. * = used in evaluation. See Table~\ref{tab:model_desc} for evaluated recordings}	
%	\vspace{-2pt}
%	\begin{tabular}{l l l l}
%		\hline
%%		\rowcolor[HTML]{a5e6a5}
%		\hlineB{2}
%		\textbf{SoC} & \textbf{GPU} & \textbf{Recorder}  & \textbf{Replayer} \\ \hlineB{2}
%		Hikey960	& Bifrost G71	& User (xxx) & User (xxx)		\\ 
%		OdroidC4	& Bifrost G31	& User (xxx) & User (xxx)		\\ 
%		OdroidN2	& Bifrost G52	& User (xxx) & User (xxx)		\\ 
%		Raspberry Pi4 & VideoCore IV& Kernel (xxx) & Kernel (xxx)	\\ \hline
%		\hlineB{2}
%	\end{tabular}
	
	\includegraphics[width=0.47\textwidth{}]{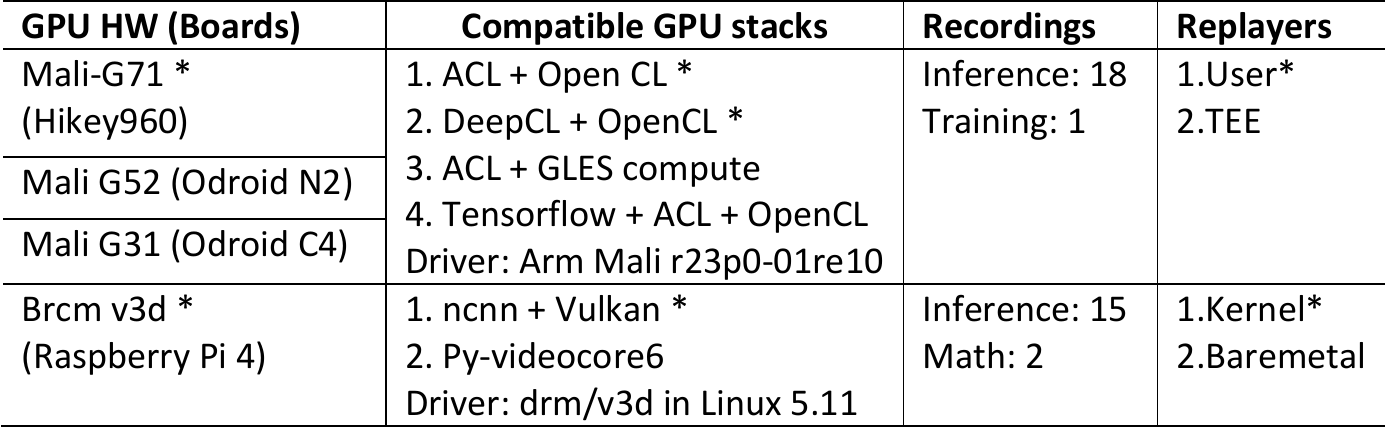} 
	
%	\vspace{3pt}		% use as needed
	\label{tab:impl}
%	\vspace{-5pt}		% use as needed	
	\end{center}
\end{table}

\section{Implementations and Experiences}
\label{sec:impl}

As summarized in Table~\ref{tab:impl}, we implement \sys{} for Arm Mali (reported to ship billions of devices~\cite{arm-mali-all}) and Broadcom v3d (the GPU for RaspberryPi 4). 
The current implementations work for a variety of ML workloads (inference, training, and math kernels), programming abstractions 
(OpenCL, Vulkan, and GLES compute), and 
GPU runtimes (the official ones as well an experimental runtime fully written in Python). 
% Besides the above stacks, we tested \sys{} on Pyvideocore6, an experimental GPU runtime fully in written Python (it implements several math kernels) and shows behaviors different from official runtimes (e.g. mapping large, contiguous GPU memories). 

% ACL does not support training. Achieve 1.4x faster learning.

\begin{comment} ======= useful ? ==========
\textbf{Individual compute kernels (shaders?)} We choose 6 simple benchmarks which has only a single job or consists of less than ten jobs.
\begin{myenumerate}
	\item \texttt{VecAdd(VA)}, 4K integer vector addition, which we implemented.
	\item \texttt{Scale(SC)}, scales up the 227x227 image.
	\item \texttt{AbsDiff(AD)}, calculates absolute difference between two images
	\item \texttt{SGEMM (MM)}, a matrix multiplication.
	\item \texttt{DirectConv(DC)}, a simple convolution without optimization.
	\item \texttt{MatrixConv(MC)}, a convolution based on matrix multiplication.
\end{myenumerate}
\end{comment}

\subsection{The Recorder for Arm Mali}
% FYI - job chain description from panfrost engineers (https://www.youtube.com/watch?v=qtt2Y7XZS3k)
%The Mali Bifrost family is our reference hardware.
%they support OpenCL 1.1/1.2/2.0 on Linux. 
%The Bifrost family of Mali is our reference hardware. 

\begin{comment}
%--- useful ---- 
%It is a family of \Note{??} mobile GPU models. 
% Arm has implemented runtimes to support OpenGL/ES, OpenCL, and Vulkan. 
%They support OpenCL 1.1/1.2/2.0 on Linux. 
Recent Mali GPUs are autonomous to support large compute jobs. 
A Mali job chain contains a job's all commands, metadata, and shader code. 
The job chain holds multiple references linking the GPU code and data. 
Mali runs a hardware scheduler to dispatch a job over its shader cores; 
it has 4-level page tables with page permissions.  
% Similar to v3d, the official Bifrost driver implements trace accessors and tracepoints and hence, we only add 500 SLoC for recording.
% The job chain size varies from types of job (e.g. compute job - 512 bytes).Mali has more hardware features so that we only put some info to the job chain then GPU does things.
% The page table in the Bifrost contains 'valid' field in the entry. So we can filtered out unnecessary memory dumps (e.g. intermediate data) and hence reduce the recording size.
\end{comment}

%\paragraph{Status}
%The Mali Bifrost family is our reference hardware.
%We implemented \sys{} for three Mali Bifrost GPUs. 
We implement a recorder for Mali Bifrost family;
it records complex and diverse GPU workloads, including 18 inferences and 1 training, some of which will be evaluated in Section~\ref{sec:eval}.
%\sys{} can record and replay complex and diverse GPU workloads, including 18 inferences and 1 training, some of which will be evaluated in Section~\ref{sec:eval}.
Leveraging ArmNN~\cite{arm-nn},
our prototype for Mali is compatible with TensorFlow NN models.
%\sys{} is compatible with ML optimization frameworks such as TVM: it can record TVM NNs, e.g. vgg16. \Note{more details?}
%https://tvm.apache.org/2018/01/16/opt-mali-gpu
% In total, we have collected recordings for over XXX NNs. 
We add around 700 SLoC to Mali's stock driver,
which is 1\% of the driver's 45K SLoC. 
%Our recorder adds around 700 SLoC to Mali's stock driver,
%which is 1\% of the driver's 45K SLoC. 
% We build the recorder oblivious to Mali's job chain format. 
%The recorder taps into a number of functions in the stock driver: 
%register accessors, system tracepoints of debugging, and performance monitoring, which is meant for Arm's DSTREAM profiler.

%\paragraph{Optimization with GPU page table semantics}
Our recorder exploits Mali's page permission to shrink memory dumps. 
% with the following heuristics. 
If a GPU-visible page is mapped as \textit{executable} to GPU, the recorder treats the page as part of job chains and dumps it. 
If a GPU-visible page is \textit{non-executable} to GPU and is \textit{unmapped} from CPU, 
the recorder treats the page as part of GPU internal buffers and excludes it from dumping. 
%Based on our analysis of the stock driver, 
This is because GPU-visible pages are mapped to CPU on demand; an unmapped page must never have been accessed by CPU. 

\subsection{The Recorder for Broadcom V3D}
%We implemented \sys{} for v3d, the GPU on Raspberry Pi 4 (Rpi4) because of its popularity. 
%The v3d GPU supports Vulkan 4.2 and OpenGL ES 3.1. 
%Compared to Mali, v3d has simpler hardware. 
%The GPU jobs include binning, rendering, compute, and texturing, all seen in the ML workloads we tested. 
%The runtime emits GPU commands in nested ``control lists''. 
% The GPU has dedicated registers for shader code addresses and control list addresses. 
% A control list may point to other lists. 
% A control list may point to other lists in the same GPU memory region. 

%\paragraph{Status}
% \sys{} works for XXX inferences of NNs from NCNN, a popular mobile deep learning framework. 
Our recorder for v3d adds around 1K SLoC to v3d's stock driver. % which is 10\% of the driver's 5K SLoC. 
% The recorder does not change job submission as the stock driver already does synchronous job submission. 
To dump GPU memory, the recorder follows v3d's registers pointing to shaders and control lists. % and dumping the entirety of enclosing memory regions. 
It handles the cases where lists/shaders may contain pointers to other lists/shaders of the same or different memory regions. 
% However, capturing only that list is insufficient as the list may refer other lists in GPU memory. 
%The stock v3d driver implements register accessors and XXX tracepoints for various GPU events, which we tap in for recording callbacks. 
Unlike Mali, the v3d page tables lack executable bits. 
Being conservative, the recorder has to dump more pages than Mali in general. 
To further exclude unwanted GPU memory regions from dumping, 
the recorder exploits as hints the flags of syscalls that allocate the GPU memory. 
To reduce the storage overhead, the recorder compresses the memory dumps with zlib~\cite{zlib}. 
% To mitigate, our recorder does some simple optimizations including compressing pages with all zeros. 
% Hence, the recorder has to include all GPU-visible pages in a dump. 
% Be conservative, the recorder has to dump more GPU pages, which results in larger dumps than those of Mali. 

%\textbf{Selective memory dump/page table} - when GPU memory mapping, the use of memory mapping can be known from device driver side. (e.g. if sync, this can be used for in/ouput). It also keeps valid field in the page table entry. So we can filtered out unnecessary memory dumps (e.g. intermediate data) as marked as no longer needed (in reality, the request comes from user-space runtime).

%3 types of jobs/ but we observed that for the inference, the stack only uses compute job type.
%We build the recorder by adding xx lines of code in official Bifrost driver (XXX SLoC).

\subsection{Replayers in Various Environments}
\label{sec:replayer-env}
%We focus on the portability of replayer implementations, which are summarized in Table~\ref{tab:impl}. 
%We implemented replayers as summarized in Table~\ref{tab:impl}. 

\paragraph{A baremetal implementation} 
As a proof of concept, 
we built a standalone replayer for v3d without any OS. 
%When the RaspberryPi 4 (Rpi4) board powers up, it directly launches the replayer. 

To avoid filesystems, we statically incorporate compressed recordings in the replayer binary. 
The whole executable binary (excluding recordings) is around 50 KB. 
In the executable, 
the replayer itself is about 8 KB. 
We link zlib~\cite{zlib} for recording decompression (about 9 KB) and a baremetal library~\cite{rpi-circle} for Rpi4. 
The library functions include
CPU booting, interrupts, exception, and firmware interfaces 
(about 15 KB executable); 
CPU cache, MMU, and page allocation (4 KB); 
timers and delays (4 KB); 
string manipulation and linked lists (9 KB). 

A major challenge is to bring up the GPU power and clocks. % which is missing in Circle.
Modern GPUs depend on power/clock domains at the SoC level~\cite{central-pm-asplos}. 
%These hardware configuration is SoC specific and complex, % e.g. the manual for XXX has XXX pages on this topic.
Linux configures power and clocks by accessing various registers, sometimes communicating with the SoC firmware~\cite{rpi-firmware}. 
The process is complex, SoC-specific, and often poorly documented. 
% For instance, the Linux has 3.3K lines of power and clock drivers for Bcm2711, the SoC for v3d. 
%While replayers atop Linux (kernel or user) reuse the configuration done by the kernel transparently, 
While replayers at the user or the kernel level reuse the configuration done by the kernel transparently, 
the baremetal replayer must configure GPU power and clocks itself. 
To do so, we instrument the Linux kernel, extract the register/firmware access, and port it to the replayer. 

\paragraph{A user-level implementation}
%We built a replayer for Mali as a user-level daemon in the spirit of kernel-bypassing drivers~\cite{dpdk,uio}. 
We built a replayer for Mali as a daemon with kernel bypassing~\cite{dpdk,uio}. 
To support the daemon, the kernel parses the device tree and exposes to the userspace the GPU registers, memory regions, and interrupts. 
The replayer maps GPU registers and memory via mmap(); 
it directly manipulates GPU page tables via mapped memory; 
it receives GPU interrupts by select() on the GPU device file.
%The replayer allocates GPU memory via \texttt{alloc\_pages} and maps the memory as DMA regions to satisfy the hardware requirement of Kirin 960 SoC.

% Old writing. Use UIO
\begin{comment}
We built a user-level daemon atop the Linux UIO framework. 
The kernel parses the device tree and exposes GPU resources to the userspace. 
The replayer maps GPU registers and GPU memory via the mmap() syscall. 
%The replayer allocates GPU memory via \texttt{alloc\_pages} and maps the memory as DMA regions to satisfy the hardware requirement of Kirin 960 SoC.
The replayer allocates GPU memory via the XXX syscalls() and requests the memory from XXX DMA regions to satisfy the hardware requirement of XXX SoC. 
The replayer receives GPU interrupts by select() on the GPU device file~\cite{dpdk,uio}.

In its current implementation, we build a tiny kernel module (700 SloC) in lieu of UIO support. The kernel modules simply maps GPU registers to the replayer, allocates GPU memory upon requests by the replayer, and delivers GPU interrupts to the replayer through a file descriptor. 
\end{comment}

\paragraph{A kernel-level implementation}
We built a replayer for v3d as a kernel module.
The replayer directly invokes many functions of the stock GPU driver, e.g. for handling GPU interrupts and memory exceptions;
it exposes several IOCTL commands for an app to load a recording and inject/extract input/output.
%The replayer module exposes several IOCTL commands for an app to load a recording and inject/extract input/output. 
Once turned on, the replayer disables the execution of the stock driver until replay completion or GPU preemption. 
% For v3d, the replayer only adds \Note{800} lines of code. 

\paragraph{A TrustZone implementation} 
We built a replayer for Mali in the secure world on the Hikey960 board. 
We added a small driver (in 100 SLoC) to the TrustZone kernel (OPTEE) for switching the mappings of GPU register and memory between the normal/secure worlds.
% The driver accepts a XXX command from the normal world OS for GPU preemption. 
The replayer is a straightforward porting of the user-level replayer. 
The replayer is in around 1K SLoC, only 0.3\% of the whole OPTEE (300K SLoC).

%\textit{Replay on TEE} \Note{move here to merge}
%Besides porting replayer to TrustZone, the MMIO lockdown is required to guarantee strong security. This can be done with TZC-400 and TZASC (\jin{ can elaborate more if needed})to annotate GPU physical memory only accessible from secure world and hence secure GPU.

% !TeX root = main.tex


% drm code : 37215 (src) + 10563 (include) = 48K

% Count cloc of MESA user space drvier (open-srouce) although it does not support OCL
% We aggregate both front and backend parkt of user space driver that might be larger than the vendor's due to IR stuffs.

\begin{table}[]
	\centering
%	\fontsize{8}{9.5}\selectfont
	
%	\begin{tabular}{r V{1} r r r r}
%%		\hline
%%		\rowcolor[HTML]{C0C0C0} 
%			\hlineB{2}
%			\textbf{GPU model} & 
%			\begin{tabular}[c]{@{}r@{}}\textbf{Runtime}\\ \textbf{binary (MB)}\end{tabular} &
%			\begin{tabular}[c]{@{}r@{}}\textbf{Driver}\\ \textbf{(SLoC)}\end{tabular} &
%			\begin{tabular}[c]{@{}r@{}}\textbf{Recorder}\\ \textbf{(SLoC)}\end{tabular} &
%			\begin{tabular}[c]{@{}r@{}}\textbf{Replayer}\\ \textbf{(SLoC)}\end{tabular} \\
%			\hlineB{2}
%			Mali Bifrost (orig)	& 48	 	& 45K 		& N/A 	& N/A	\\ %\hline
%			Bcrm v3d (orig)		& XXX		& 3K		& N/A 	& N/A	\\ %\hline
%			NV Tegra X1/K1		& 16		& 214K 		& N/A 	& N/A	\\ 
%			Qualcomm Adreno		& XXX		& XXX		& N/A	& N/A	\\
%			\hline \hline
%			GPURip (Mali)		& 0.03		& 0.7K 		& 0.7K 	& 2.2K	\\ %\hline
%			GPURip (v3d)		& XXX		& XXX 		& XXX 	& XXX	\\ %\hline
%			
%%			\textbf{Imagination PowerVR}	& N/A 		& 66K 		& (66 + $\alpha$)K      \\ %\hline
%			\hlineB{2}
%	\end{tabular}
	\caption{Codebase comparisons. Binaries are stripped.}
	
	\includegraphics[width=0.47\textwidth{}]{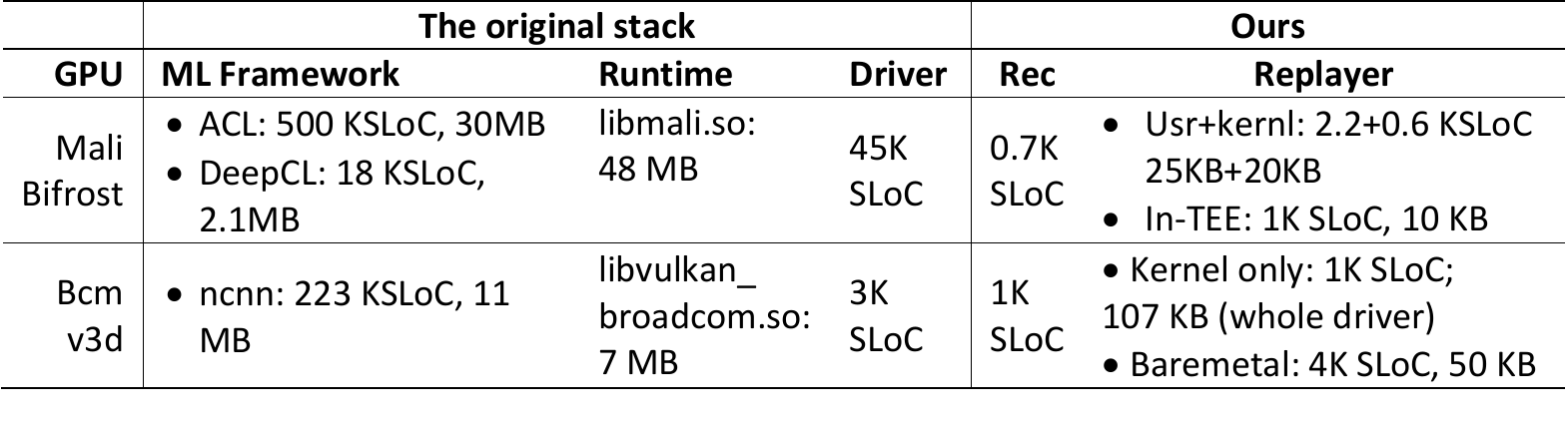} 
	
%	\vspace{-6pt}		% use as needed
	
	\label{tab:src-gpu-stack}
%	\vspace{-20pt}		% use as needed

\end{table}

\subsection{Reusing Recordings Across GPU SKUs}
\label{sec:reusing}
%In general, a GPU recording is only for a specific GPU hardware model, as the recording contains register info, etc. 
%Our experiences suggest it is possible to record/replay on different GPUs of the same family. 
It is possible to share recordings across GPUs of the same family: 
these GPUs are likely to share job formats, shader instruction sets, and most register/page table semantics.
%We examine that the GPU models of the same family (e.g. Mali Bifrost) share the same register map and device driver;
%a consolidated driver supports 6 (Bifrost family) and 17 (adreno 6xx) GPUs, handling each GPU model's quirks based on a set of register values (e.g. XXX regs).
%We also observe the case that GPUs of different families (Mali Midgard and Bifrost) share the almost same register map.
%The discrepancies are often minor, and can be fixed through patching. 
% G31 efficient, G52 mainstream, G71 high end
%We analyze the register traces and memory dumps of three Mali GPUs: 
We analyze three Mali GPUs: 
G31 (low end), G52 (mainstream), and G71 (high end).
We manage to patch a recording from G31/G52 and replay it on G71. 
Our patch adjusts: 
(1) Page table format: re-arranging the permission bits in the G31 page table entries, which are in a different order than G71 due to G31's LPAE support. 
% Whereas it is not necessary for G52 recording as it shares same page table format with G71.
(2) MMU configuration: 
%in G31/G52 recordings, 
flipping a bit in the translation configuration register 
%(AS\_TRANSCFG) 
to enable read-allocation caching expected by G71. % but unsupported in G31/52.
(3) Core scheduling hints: 
%in G31/G52 recordings, 
changing the value of core affinity register (JS\_AFFINITY) so a job is mapped to G71's all 8 shader cores. 
% from 0x1 and 0x3 to 0xFF, respectively, as G71 has 8 shader cores.
% \Note{G31-2 cores, G52-4 cores}
%  Note that the bitmask represents the number of cores. 
%A patch is small: (1) and (2) update one register write per recording and (3) requires one register modification per job.
Overall, the patch includes fixes for two registers per recording and one register per job.
%A patch is as few as XXX bytes per recording. 
Section~\ref{sec:eval-design} reports replay performance of a patched recording. 

Despite our limited success above, 
we note that it would be difficult to replay with fewer GPU resources (e.g. record on G71 and replay on G31). 
This is because doing so would require (1) proprietary GPU knowledge, e.g. to relocate GPU shaders and compact memory and (2) a more sophisticated replayer, e.g. to swap GPU memory. 

\begin{table*}[]
	\centering
	\vspace{1pt}
	\fontsize{6.7}{9.5}\selectfont
	\caption{\sys{} eliminates common vulnerabilities and exposures (CVEs) in the GPU stack}
	\begin{tabular}{|l|l|l|l|l|}
		\hlineB{1.5}
		\cellcolor{mylightgray} \textbf{\sys{}'s design (D1--3: scenarios)} &
		\cellcolor{mylightgray} \textbf{Example CVEs} &
		\cellcolor{mylightgray} \textbf{Description} &
		\cellcolor{mylightgray} \textbf{Effect} &
		\cellcolor{mylightgray} \textbf{Vulnerability} \\

		% By design, we only run 1 replayer at a time.
		\hline \hline
		\multirow{3}{*}{\begin{tabular}[c]{@{}l@{}}\textbf{Remove GPU runtime from app}\\ \textbf{(D1,D2,D3)}\end{tabular}}
		& CVE-2014-1376, High	& Improper restriction of OpenCL calls~\cite{CVE-2014-1376}				& Arbitrary code execution			& \texttt{App.I} \\ \cline{2-5}
		& CVE-2019-5068, Med    & Exploitable shared memory permissions~\cite{CVE-2019-5068} 	      	& Unauthorized mem access 			& \texttt{App.C} \\ \cline{2-5}
		& CVE-2018-6253, Med    & Malformed shaders cause infinite recursion~\cite{CVE-2018-6253} 		& App hang                        	& \texttt{App.A/GPU.A} \\ \hline \hline
		
		\multirow{5}{*}{\begin{tabular}[c]{@{}l@{}}\textbf{Remove GPU driver }\\ \textbf{(D2, D3) }\end{tabular}}
		& CVE-2017-18643, High 	& Leak of GPU context address of GPU mem region~\cite{CVE-2017-18643} 	& Sensitive info disclosure			& \texttt{Kernel.C} \\ \cline{2-5}
		& CVE-2019-20577, High 	& Invalid address mapping of GPU buffer~\cite{CVE-2019-20577} 			& Kernel crash						& \texttt{Kernel.I} \\ \cline{2-5}
		& CVE-2020-11179, High 	& Race condition by overwriting ring buffer~\cite{CVE-2020-11179} 		& Arbitrary kernel mem r/w			& \texttt{Kernel.I} \\ \cline{2-5}
		& CVE-2019-10520, Med 	& Continuous GPU mem allocating via IOCTL~\cite{CVE-2019-10520} 		& GPU mem exhausted					& \texttt{Kernel.A} \\ \cline{2-5}
		& CVE-2014-0972, N/A 	& Lack of write protection for IOMMU page table~\cite{CVE-2014-0972}	& Kernel mem corruption				& \texttt{Kernel.I} \\ \hline \hline
		
		\textbf{\begin{tabular}[c]{@{}l@{}}\textbf{Disable fine-grained GPU sharing} (\textbf{D1,D2}) \end{tabular}} % no fine-graained GPU sharing
		& CVE-2019-14615, Med	& Learning app's secret from GPU register file~\cite{CVE-2019-14615} 	& App data leak & \texttt{App.C} \\ \hline
	\end{tabular}

	\texttt{I}: Integrity;
	\texttt{C}: Confidentiality;
	\texttt{A}: Availability
	
%	\vspace{-1pt}
%	\caption{Example security vulnerabilities eliminated by \sys{}}
	\label{tab:security}
%	\vspace{-10pt}		% use as needed
\end{table*}

\section{Evaluation}
\label{sec:eval}

We evaluate \sys{} with the following questions.

\begin{myitemize}
	\item
	Does \sys{} make GPU computations more secure? 

	\item Overhead: Do recordings increase app sizes? How does the replay speed compared to that of the original GPU stack? 
	
	\item Do our key design choices matter? 
		
\end{myitemize}

\subsection{Analysis}
\label{sec:eval-security}

%We analyze how \sys{} mitigates attacks against a GPU stack. 

%We consider \textit{semantic bugs} in the GPU stack out of scope, which may result in flawed recording and flawed GPU execution. 
%\sys{} neither mitigates nor exacerbates these bugs. 
\noindent
\textbf{Semantic bugs}, e.g. emission of wrong GPU commands, may preexist in the GPU stack for recording. % or GPU binaries computing wrong results. 
% Such bugs may result in defected recordings, which lead to incorrect GPU results on the target machines. 
Such bugs may propagate to the target machines, resulting in wrong replay results. 
\sys{} neither mitigates nor exacerbates these bugs. 
Fortunately, semantic bugs are rare in production GPU stacks to our knowledge.
\sys{}'s recorder and replayer may introduce semantic bugs. 
%The chance, however, is slim: they are small as a few K SLoC 
The chance, however, is slim: as shown in Table~\ref{tab:src-gpu-stack}, they are small as a few K SLoC 
%(see Table~\ref{tab:src-gpu-stack}) 
with simple logic.
%The chance, however, is slim: the implementations are small as a few K SLoC (see Table~\ref{tab:src-gpu-stack}) with simple logic.
Our validation experiments in \sect{eval-replay-correctness} strengthen our confidence.
% Note that this does not include \textit{semantic bugs} in the GPU stack, which may result in recording defects and incorrect GPU execution results. 
% \sys{} neither mitigates nor exacerbates these bugs. 
We next focus on security, a major objective of \sys{}. 

\paragraph{Threat models}
%We consider threat models corresponding to three deployment scenarios in Figure~\ref{fig:deployment}: 
Corresponding to three deployment scenarios (D1-3) in \sect{intro}: 
(D1) a user/kernel-level replayer on a commodity OS trusts the OS while facing local unprivileged and remote adversaries; 
(D2) a replayer in TEE trusts the TEE kernel while facing the local OS adversaries and remote ones; 
(D3) a baremetal replayer only faces remote adversaries. 

We assume it is difficult to compromise the recording environment, including OS, GPU stack, and code signing:
doing so often requires long campaigns to infiltrate the developers' network where risk management is likely rigorous~\cite{nist-supplychain}. 
We will nevertheless discuss the consequences of such attacks.

\paragraph{Thwarted attacks}
corresponding to three deployment scenarios are as follows.
% Table~\ref{tab:security} summarizes vulnerabilities eliminated for the three deployment scenarios. 
(D1) 
When a replayer coexists with the GPU stack on the same OS, 
the app using the replayer is free of GPU runtime vulnerabilities which cause unauthorized access to app memory~\cite{CVE-2019-5068}, arbitrary code execution in the app~\cite{CVE-2014-1376}, and app hang~\cite{CVE-2018-6253}.
% it improves the target app's availability and integrity because the app is no longer linked to a GPU runtime. 
%In these scenarios, 
% --- useful ---- 
%(4) \sys{} reduces app data leaks because it disallows fine-grained GPU sharing~\cite{pixelVault,cudaLeak,lee14sp}. 
% (3) When the replayer run in the TEE, the above security properties are strengthened against a privileged OS attacker. 
%Enclosing the app and the replayer in a TrustZone TEE strengthens the above security properties since the OS is no longer part of the TCB.
% (3) A baremetal replayer eliminates the OS and further shrinks the TCB. 
(D2) When a replayer runs in TEE and coexists with the GPU stack outside the TEE, the app is free from attacks against the GPU stack by the local OS. 
% excludes the local OS from the is no longer part of the TCB.
(D3) 
When a replayer completely replaces the GPU stack in a system,
the system is free from GPU stack vulnerabilities that cause kernel information disclosure~\cite{CVE-2017-18643}, kernel crash~\cite{CVE-2019-20577}, and kernel memory corruption~\cite{CVE-2014-0972}. 
Table~\ref{tab:security} summarizes the eliminated vulnerabilities.

\paragraph{Attacks against \sys{}} % are difficult but still possible. 
% Attacks are made more difficult but still possible. 
%\textit{Attacks against recording integrity.}
(1) \textit{Attacks against developers' machines or recording distribution.}
%Remote adversaries may compromise the developers' GPU stack or tamper with the recording's transmission.  
This is difficult as described above. 
Nevertheless, successful adversaries may fabricate recordings containing arbitrary actions and memory dumps. 
A fabricated recording may hang GPU but cannot break security guarantees  enforced by the replayer, e.g. no illegal register access 
(\S\ref{sec:verficiation}).
%This is much more difficult than compromising target devices exposed in deployment.
%
%The injected GPU code can hang GPU but cannot make unauthorized memory/register accesses or consume excessive physical memory, which is prevented by the replayer (\S\ref{sec:verficiation}).
%
%Remote adversaries may compromise the developer's machine;
%this is much more difficult than compromising target devices exposed in deployment.
%Remote adversaries may corrupt a recording during its transmission.
%Successful adversaries may execute arbitrary GPU code. 
%The code may hang GPU; 
%it cannot make unauthorized memory/register accesses or consume excessive physical memory, which is prevented by the recorder (\S\ref{sec:verficiation}).
% 
(2) \textit{Attacks against the replayer or its TCB.}
The chance of replayer vulnerabilities is slim due to simplicity.
%and promise of formal verification~\cite{klein2009os}.
% the absence of blackbox runtime and the lower complexity of replayer can facilitate formal verification for further security analysis~\cite{klein2009os}.
%thanks to the absent of blackbox runtime and its lower complexity, the replayer is also amenable to formal verification~\cite{}.
%The replayer is amenable to formal verification. \Note{cite} \Note{check}. 
Nevertheless, successful adversaries may subvert recording verification. 
% 's security mechanisms, e.g. recording verification. 
% circumvent the recording verification. % security checks. 
By compromising a user-level replayer or kernel-level/baremetal replayers, adversaries may gain unrestricted access to the GPU or the whole machine, respectively.
\begin{table}[t!]
	\centering
	\caption{NN inference for evaluation. Choices of NNs for Mali vs. v3d are slightly different because their ML frameworks do not implement exactly the same set of NNs}
	
	\includegraphics[width=0.43\textwidth{}]{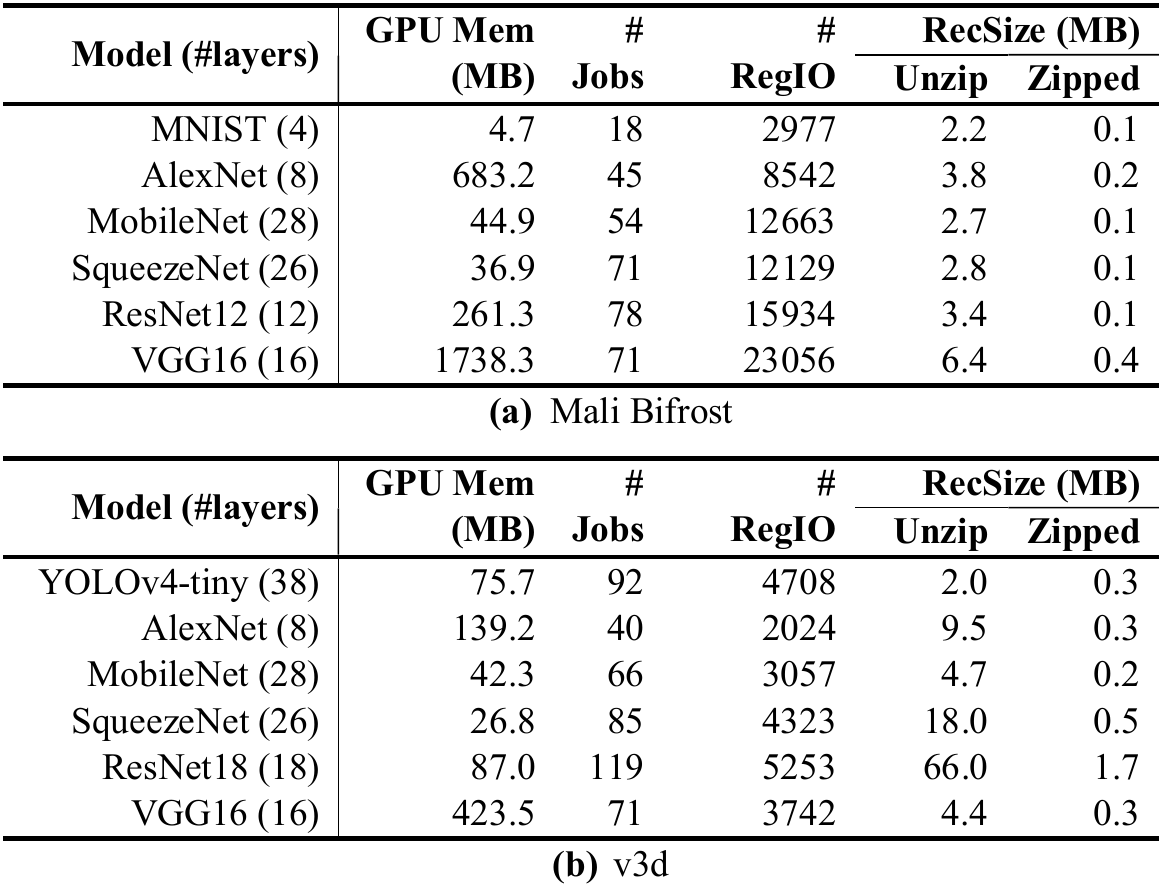} 
%	\vspace{-5pt}		% use as needed
	\label{tab:model_desc}
	\vspace{-15pt}		% use as needed
\end{table}

% !TeX root = main.tex

\begin{figure*}
	\centering
	\includegraphics[width=0.5\textwidth{}]{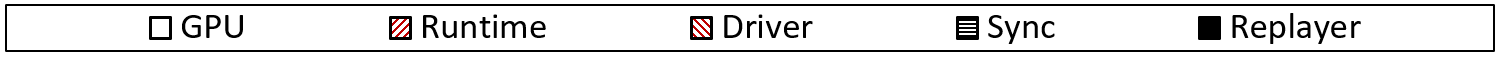}
	\vspace{-6pt}

%%%%%%%%%%%%% startup delay
	\subfloat[Mali G71]{
		\includegraphics[width=0.49\textwidth]{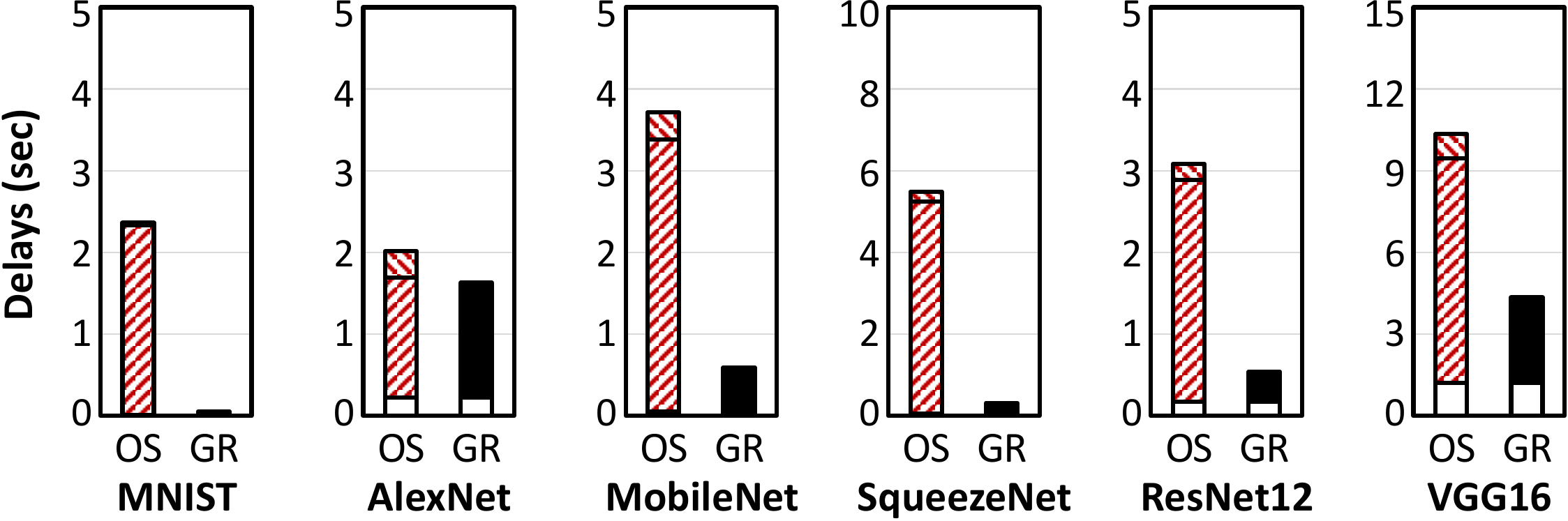}
	}
	\hfill
	\subfloat[v3d]{
		\includegraphics[width=0.49\textwidth]{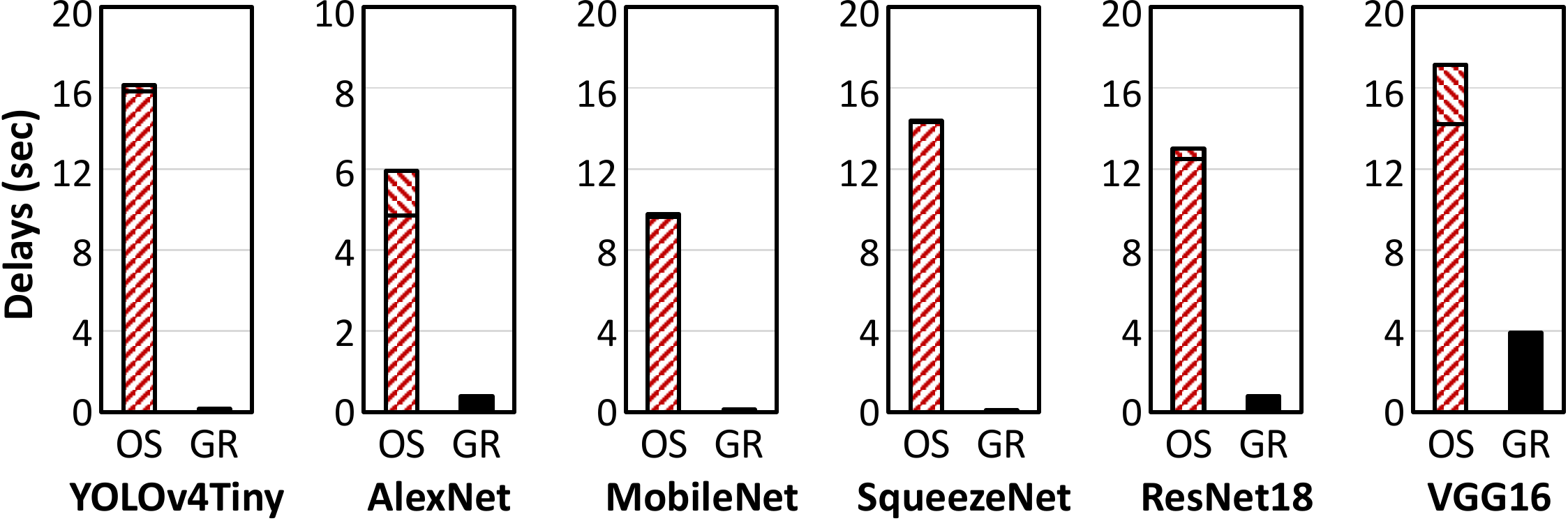}
	}
	\vspace{-7pt}
	\caption{Startup delays prior to NN inference. The replayer (GR) takes much less time than the original GPU stack (OS).}
	\label{fig:perf-ml-startup}
%	\vspace{-6pt}

%%%%%%%%%%%%% inference delay
	\subfloat[Mali G71]{
		\includegraphics[width=0.49\textwidth]{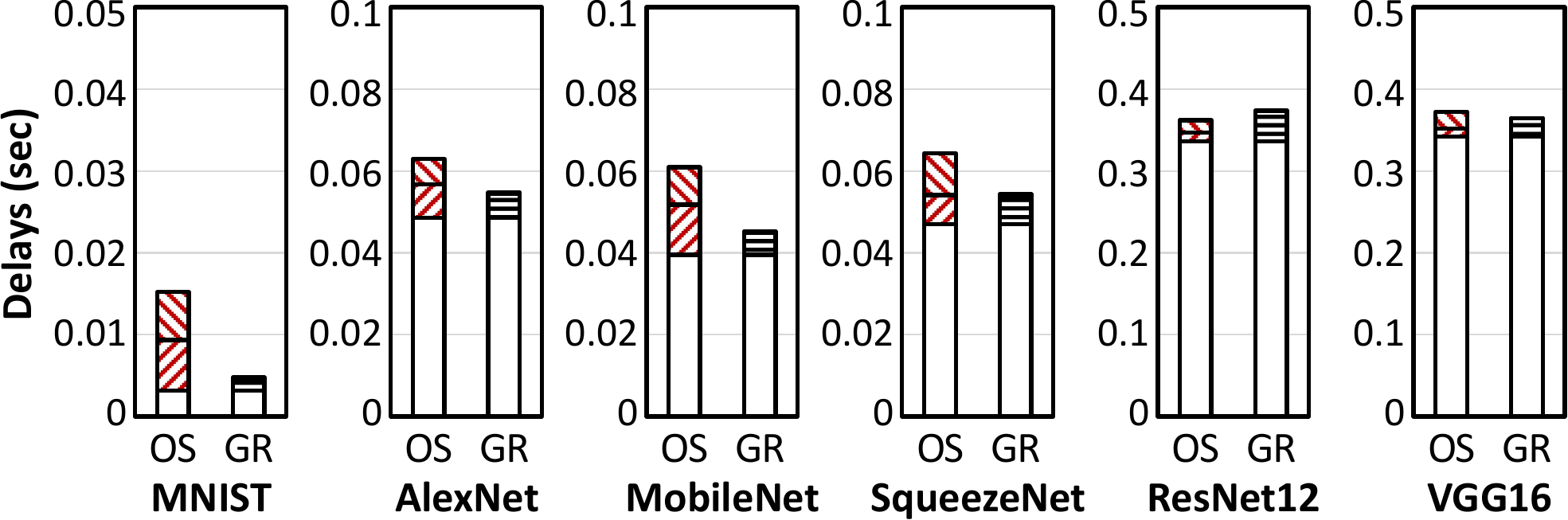}
	}
	\hfill
	\subfloat[v3d]{
		\includegraphics[width=0.49\textwidth]{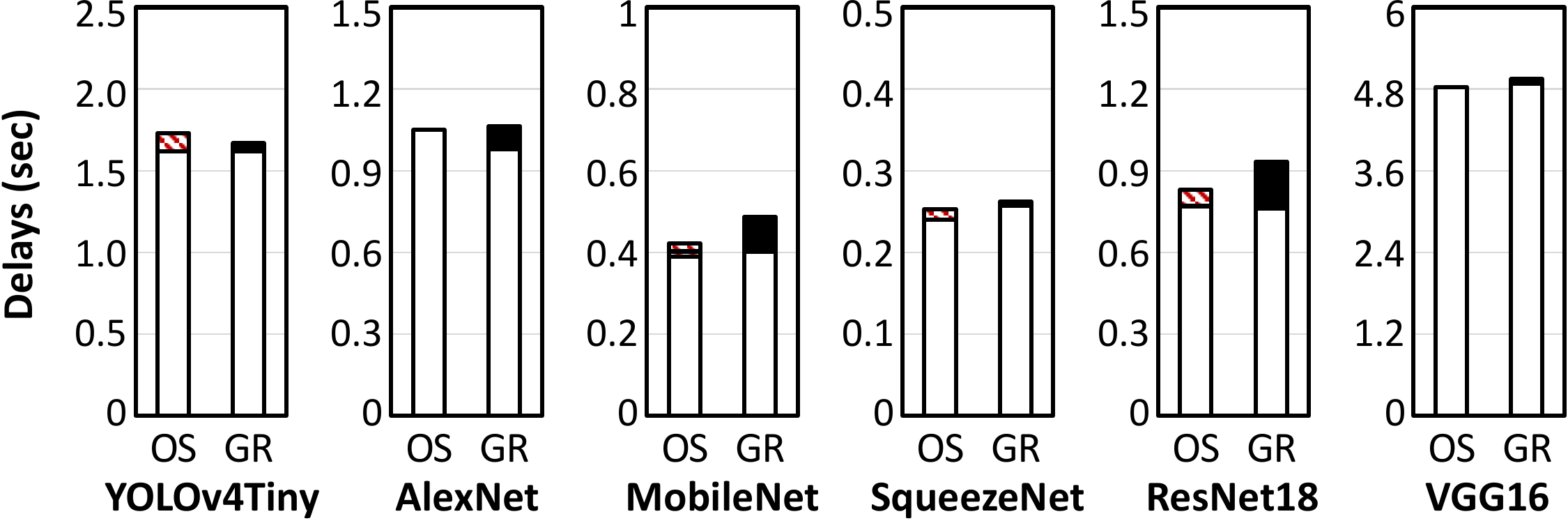}
	}
	\vspace{-7pt}
	\caption{NN inference delays. The replayer (GR) incurs similar delays as compared to the original GPU stack (OS).}
	\label{fig:perf-ml-infer}
%	\vspace{-2pt}
\end{figure*}

\subsection{Validation of Replay Correctness}
\label{sec:eval-replay-correctness}
%We experimentally verify replay correctness.
We add extensive logging to both the original driver code and the replayer:
they log \textit{all} the GPU registers on each CPU/GPU interaction;
they take snapshots of GPU memory before each job submission and after each interrupt.
We then compare these logs across runs and look for any discrepancies.

We run two inference workloads, MNIST and AlexNet, each for 1,000 times.
In each replay run, we create strong interferences with GPU by co-executing CPU programs that:
(1) generate high memory traffic which contends with GPU register and memory access; 
(2) burn CPU cycles to trigger SoC thermal throttling.
We also repeat the tests with GPU running at different clockrates. 
%From the logs (3K register and from MNIST, 8K register and 120 memory snapshots from AlexNet), 
Each MNIST (AlexNet) run generates a log of 3K (8K) registers accesses and 46 (120) memory snapshots, respectively. 
The only detected discrepancies are the numbers of register polling and GPU job delays, which do not affect GPU states;
all other logs match.
%Each MNIST (AlexNet) run generates a log of 3K (8K) register accesses and 46 (120) memory snapshots, respectively. 
%From the logs, 
%the only discrepancies between record and replay runs are the numbers of register polling and GPU job delays, which do not change GPU states. 
%the nondeterministic behaviors that does not change GPU states, which are known to the replayer already.
%Examples are register polls and variation in GPU job delays.
%All other logs match and no errors are reported.

We further verify that the replayer produces correct compute results.
% on more workloads. 
We replay all the workloads in Table~\ref{tab:model_desc} (a) 2,000 times each.
We create random input, inject interference, and compare the GPU's outcome with the reference answers computed by CPU.
The replayer always gives the correct results.
The reasons are (1) our design enforces determinism, e.g. by disallowing concurrent kernels and (2) no hardware errors during our benchmarks.

\paragraph{Failure detection \& recovery}
%To further verify the replayer's robustness to divergence, 
We run a CPU program to artificially inject 
%non-preventable state divergence 
transient, non-preventable failures during the replay of AlexNet: 
(1) offlining GPU cores forcibly and 
(2) corrupting GPU page table entries.
The replayer successfully detects the failures as diverging reads of a status register and GPU memory exceptions, because the original driver checks the register and enables the interrupt.
Re-execution resets GPU cores and re-populates the page table, finishing the execution. 

\begin{figure}
	\centering
	\vspace{-8pt}
	\subfloat[Startup prior to training]{
		\includegraphics[width=0.22\textwidth]{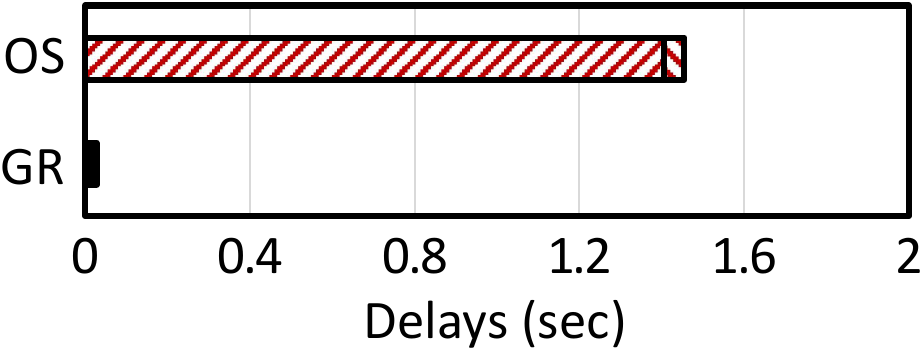}
	}
	\hfill
	\subfloat[Training (20 iterations)]{
		\includegraphics[width=0.22\textwidth]{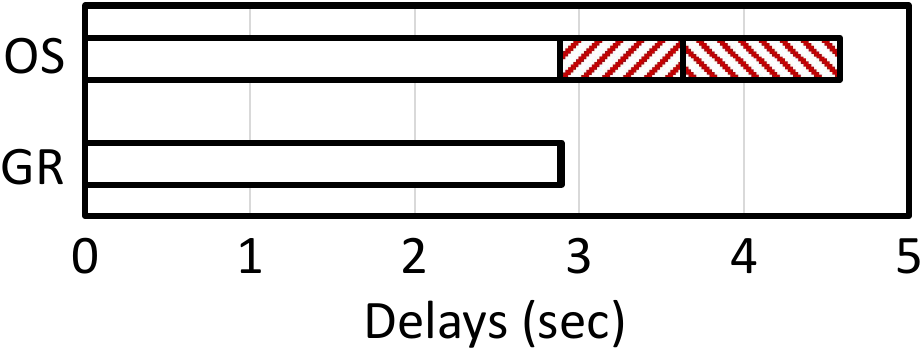}
	}
	\vspace{-2pt}
	\caption{NN training delays. Benchmark: MNIST training atop DeepCL + OpenCL on Mali G71 (OS: orig stack; GR: \sys{}).}
	\label{fig:training}
	\vspace{-8pt}
\end{figure}

% 07/30/2021 heejin
% For ASPLOS template, subfigure doesn't work
\begin{comment}
\begin{figure}
	\centering
	\begin{subfigure}[b]{0.21\textwidth}
		\includegraphics[width=\textwidth]{./figs/mnist-training-startup}
		\vspace{-16pt}
		\caption{Startup prior to training}
	\end{subfigure}
	~
	\hspace{-1pt}
	\begin{subfigure}[b]{0.21\textwidth}
		\includegraphics[width=\textwidth]{./figs/mnist-training-training}
		\vspace{-16pt}
		\caption{Training (20 iterations)}
	\end{subfigure}
	\vspace{-10pt}		% use as needed
	\caption{NN training delays. Benchmark: MNIST training atop DeepCL~\cite{deepCL} + OpenCL on Mali G71}
	\label{fig:training}
	\vspace{-10pt}		% use as needed
\end{figure}
\end{comment}
\subsection{Memory Overheads}

\paragraph{Recording sizes}
A GPU recording is as small as a few hundred KBs when compressed as shown in Table~\ref{tab:model_desc}.
% ranging from 2.2 to 6.4 MB each for Mali and from 2 to 66 MB for v3d, as shown in Table~\ref{tab:model_desc}. 
The size is a small fraction of a smartphone app, which is often tens of MBs~\cite{app-size}. 
Of a recording, memory dumps are dominant, e.g. on average 72\% for Mali. % which contains job binaries. 
%The memory dumps would have been GBs without our selective capturing (\S\ref{subsec:locating_inout}). 
Some v3d recordings are as large as tens of MBs uncompressed
because they contain memory regions that the recorder cannot safely rule out from dumping. 
Yet, these memory regions are likely GPU's internal buffers;  
they contain numerous zeros and are highly compressible. 

%\paragraph{Memory consumption}
\paragraph{CPU/GPU memory}
% We measure both CPU and GPU memory consumptions during the executions of NN inference. 
The replayer's GPU memory consumptions show a negligible difference compared to that of the original GPU stack, because the replayer maps all the GPU memory as the latter does.
%Their GPU memory consumptions show a negligible difference because the replayer does all the GPU memory mapping as the original stack.
The replayer's CPU memory consumption ranges from 2 -- 10 MB (average 5 MB) when executing NN inference, much lower than the original stack (220 -- 310 MB, average 270 MB).
% Despite the replayer lacks sophisticated memory management, 
This is because the replayer runs a much smaller codebase; 
by directly loading GPU memory dumps, it avoids the major memory consumers such as GPU contexts, NN optimizations, and JIT commands/shader generation. 

% for the original stack and \sys{}'s replayer with 6 ACL NN inference listed in Table~\ref{tab:model_desc}.

%We have measured GPU memory consumptions for the original stack and \sys{}'s replayer and cannot notice any difference. 
%This is because the replayer does all the GPU memory mapping as the original stack. 
%The replayer's CPU memory consumption (2-10 MBs) is much less than that of the the original stack (220-310 MBs). 
%This is because the replayer runs a much smaller codebase and avoids major memory consumers such as JIT generation of shaders and GPU commands. 

% !TeX root = main.tex

\begin{figure}
	\centering
	\includegraphics[width=0.41\textwidth{}]{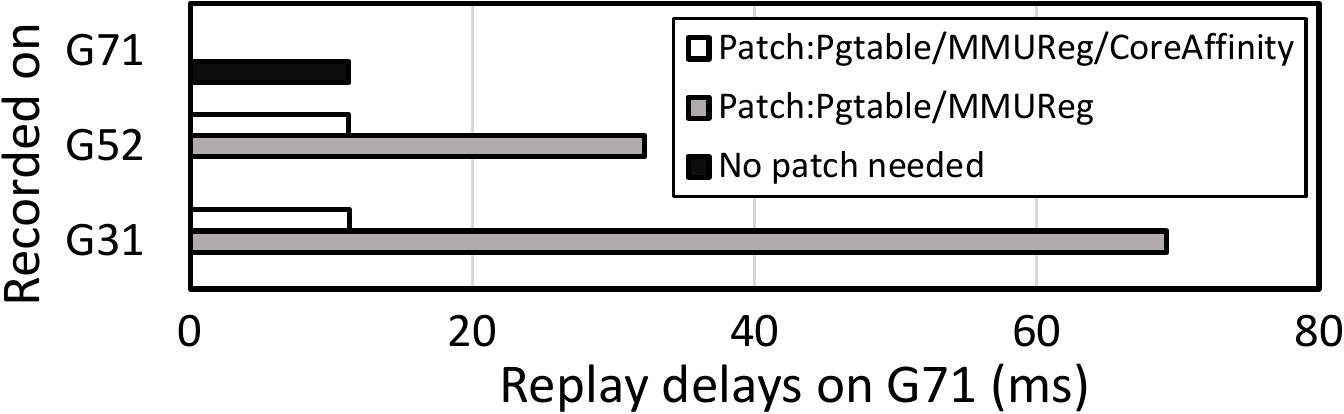}
	\vspace{-6pt}		% use as needed
%	\caption{Record 16M vector addition from Mali G31/G52 and replay them with on G71. After patching, the \sys{} can run both traces utilizing all the cores of G71. Compute: vecadd in OpenCL and ACL.}
%	\caption{Mali G71 can replay recordings from other GPUs at full hardware speed. Benchmark: vecadd from ACL~\cite{acl}}
	\caption{Mali G71 can replay recordings from other GPUs at full hardware speed. Benchmark: 16M elements \texttt{vecadd}}
		
%		with the comparable performance 
%		Patching both traces, the \sys{} can run both trace with the comparable performance utilizing all the cores .
%	Record 16M vector addition from Odroid N2 (Mali G52) and replay it in the Hikey960 (Mali G71). 
%	G71->G71: both record and replay on G71. 
%	G52->G71 (2c): patch all fields but not the XXX which only uses G71's 2 shader cores (as a reference point). 
%	G52->G71 (8c): patch all fields including XXX to use G71's all 8 shader cores.
	\label{fig:reusability}
	\vspace{-8pt}		% use as needed
\end{figure}

%\subsection{Replay delays}
\subsection{Replay Speed}
\label{sec:eval-perf}

We study the inference delays on a variety of NNs as listed in Table~\ref{tab:model_desc}.
Compared to the original GPU stacks (native execution), 
the replayer's startup delays are significantly lower: by 26\% -- 98\% (Mali) and lower by 77\% -- 99\% (v3d);   % 0.1 -- 15\%
Our replay is even 20\% faster (Mali) and only 5\% slower (v3d) on average. 
Our overhead is much lower than prior TEE systems for secure GPU computation~\cite{HIX,slalom,graviton}.
% e.g. 150% in HIX [46]; 800% in Slalom [98]; 30% in Graviton [99]. 

% --- old --- 
%the replayer's execution delays are similar: ranging from 44\% lower to 13\% higher (Mali) and from 4\% lower to 19\% higher (v3d). 

\paragraph{Startup delays}
We measure the startup delay from 
the time the testing app initializing a GPU context
until the first GPU job is ready for submission. 
Figure~\ref{fig:perf-ml-startup} shows the results.  
Both the stacks for Mali and v3d take seconds to start up, yet showing different bottlenecks: 
Mali is bottlenecked at the runtime (libMali.so) compiling shaders and allocating memory; v3d is at the framework (ncnn) loading NNs and optimizing pipelines. 
By contrast, the replayer spends most time on GPU reset, loading of memory dumps, and reconstructing page tables. 
% By contrast, \sys{} is much faster as it simply replays the final state of initialization (e.g. shader code and memory mappings). 

% The startup comparison should \textit{not} lead to quantitative conclusion, though.
Our startup comparison should \textit{not} be interpreted as a quantitative conclusion, though.
We are aware of optimizations to mitigate bottlenecks in GPU startup, 
e.g. caching compiled shaders~\cite{mali-offcompile} or built NN pipelines~\cite{ncnn-pipelinecache}. 
Compared to these point solutions,
\sys{} is systematic and pushes the caching idea to its extreme -- caching the whole initialization outcome at the lowest software layer. 

%By capturing all the data after startup is completed, the \sys{} can bypass such process and achieve the same thing by simply loading the trace.
%For instance, the \sys{} eliminates user delay mostly results from JIT compilation and hence shares a benefit of offline GPU compilation~\jin{cite}; it also squeezes the kernel delay that comes from scheduling and memory allocation for the computation.
%The \sys{}'s delay is negligible which includes loading data of job binaries and reconstructing GPU page table;
%besides compute overhead, it is only compared to the full stack.

% When running inference, the GPU stack should process extra operations such as preparing GPU code by JIT compilation, tuning the kernel, fusing the layers for graph optimization, and running some kernels for weight and parameter reshaping.

\paragraph{NN inference delays}
We measured the delay from the moment an app starting an inference with its ML framework to the moment app getting the outcome. 
The results are shown in Figure~\ref{fig:perf-ml-infer}. 
In general, on benchmarks where the CPU overhead is significant, the replayer sees lower delay than the full stack, e.g. by 70\% on MNIST (Mali). 
This is because the replayer minimizes user-level executions, kernel-level memory management, and user/kernel crossings such as IOCTLs. 
On larger NNs with long GPU computation, \sys{} sees diminishing advantages and sometimes disadvantages. 
% e.g. around 3\% and 12\% longer on ResNet12 (Mali) and ResNet18 (v3d).
%e.g. around 13\% longer delays on AlexNet (Mali), and ResNet18 (v3d). 
%Examining these benchmarks, 
\sys{}'s major overheads are (1) loading of memory dumps containing unneeded data that \sys{} cannot exclude, e.g. 66 MBs for ResNet18 (v3d); 
% \jin{need discuss: load dump is done when startup on Mali}
(2) short GPU idles from synchronous jobs (0.5\% -- 3\% on Mali); 
(3) pause between replay actions. 
%(3) pause intervals that \sys{} cannot safely remove. % (others). 
% \sys{}'s overhead is higher on v3d, primarily because of the memory dumps are larger and hence take longer to load. 

\paragraph{NN training delays}
\sys{} shows similar advantages. % on training. 
Our benchmark is MNIST with DeepCL~\cite{deepCL} atop OpenCL.
Each training iteration runs 72 GPU jobs and 5.7K register accesses.
DeepCL already submits jobs synchronously with CLFlush().
As shown in Figure~\ref{fig:training}, the replayer incurs 99\% less startup delay due to the removal of parameter parsing and shader compilation.
% Note that unlike ACL, the DeepCL does not have GPU compute (e.g parameter reshape) in startup time.
Over 20 iterations, the replayer incurs 40\% less delays because it avoids DeepCL and the OpenCL runtime.

\begin{figure}
	\centering
	\includegraphics[width=0.4\textwidth{}]{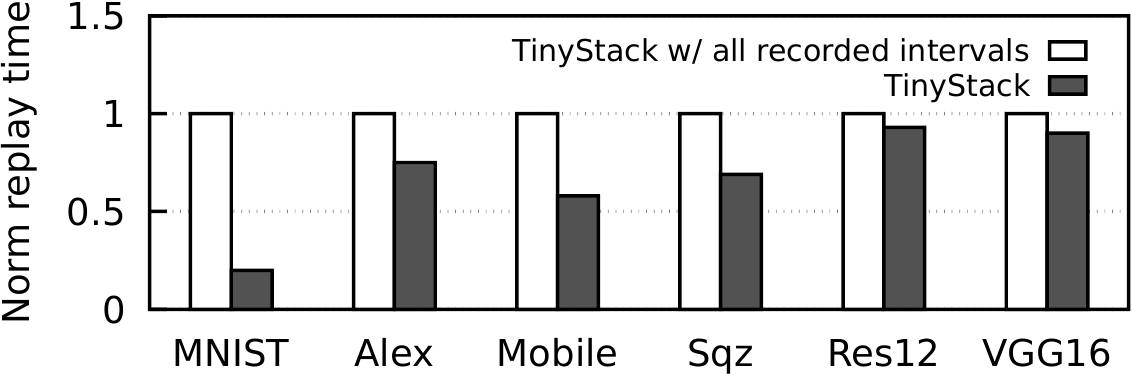}
	\vspace{-5pt}		% use as needed
%	\caption{\sys{} removes unnecessary intervals between replay actions to speed up replay. Benchmark: NN inference atop Arm Compute Library and Mali G71}
	\caption{\sys{} removes unnecessary intervals between replay actions. Benchmark: ACL NN inference atop Mali G71}
	\label{fig:interval}
	\vspace{-4pt}		% use as needed
\end{figure}

\subsection{Validation of Key Designs}
\label{sec:eval-design}

\paragraph{Cross-GPU record/replay} (\S\ref{sec:reusing})
%We demonstrate the feasibility of record and replay on different GPUs of the same family (\S\ref{sec:reusing}).
Figure~\ref{fig:reusability} demonstrates it on different GPUs of the same family.
We have recorded the same workload on Arm Mali G31 (low-end, 1 shader core) and G52 (mainstream, 2 cores). We attempt to replay the two recordings on Mali G71 (high-end, 8 cores).
With patched GPU page tables and MMU register values, the replay completes with correct results, albeit with 4x -- 8x lower performance. 
Further patching the core affinity register makes the replay utilize G71's all 8 shader cores, resulting in full performance.

%Figure~\ref{fig:reusability} shows the feasibility of replaying on a different GPU model. 
%The two recordings are produced on Arm Mali G31 and G52. 
%After manually patching the recorded page table entries and MMU register values, we are able to replay the recordings on Mali G71. 
%The results show replaying the patched recording is as fast as the recording captured on the same GPU (G71). 
%Our experiment demonstrates feasibility of the cross compatibility among different GPUs with same architecture.

\paragraph{Skip intervals in replay} (\S\ref{subsec:delay_reduction})
%The recorder's removal of intervals is significant to replay speed. 
Without the technique, the replayer's NN inference will be 1.1x -- 4.9x longer, as shown in Figure~\ref{fig:interval}; 
startup delays will be up to two orders of magnitude longer, closer to that of a full stack (not shown in the figure). 
%This is because the recorder can no longer recognize the intervals of CPU busy execution caused by runtime and exclude them from the recording. 

\begin{figure}
	\centering
	\vspace{-8pt}
	\subfloat{
		\includegraphics[width=0.135\textwidth]{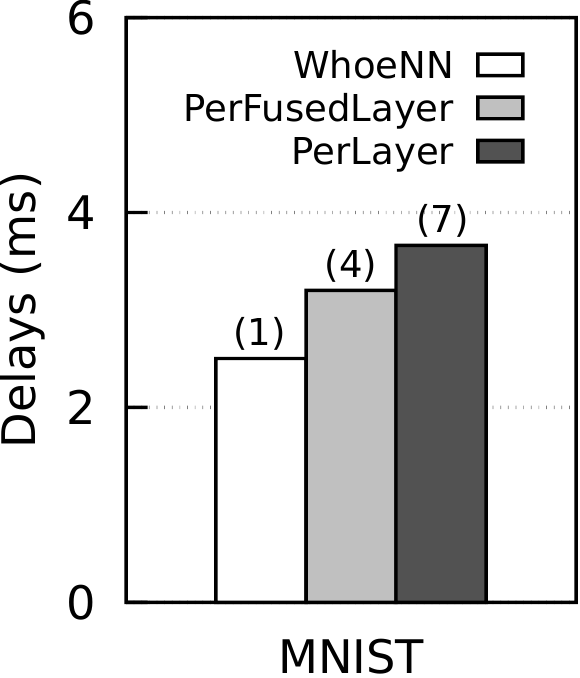}
	}
	\hspace{-2pt}
	\subfloat{
		\includegraphics[width=0.145\textwidth]{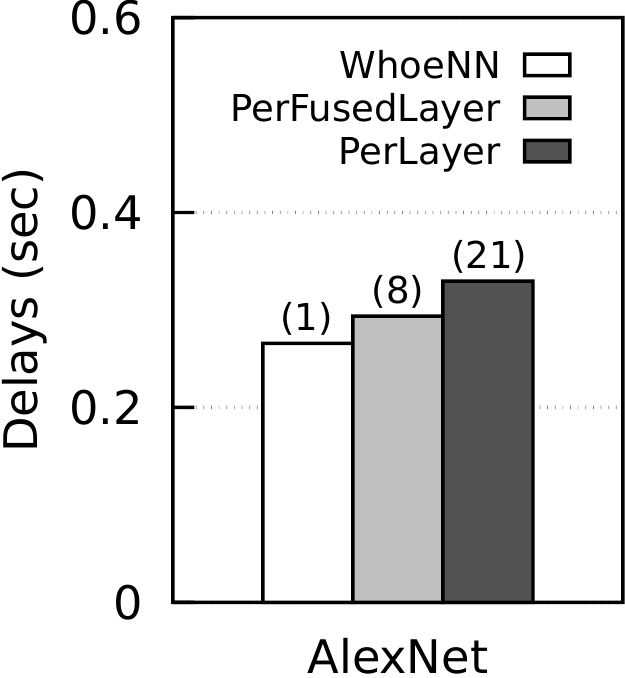}
	}
	\subfloat{
	\includegraphics[width=0.145\textwidth]{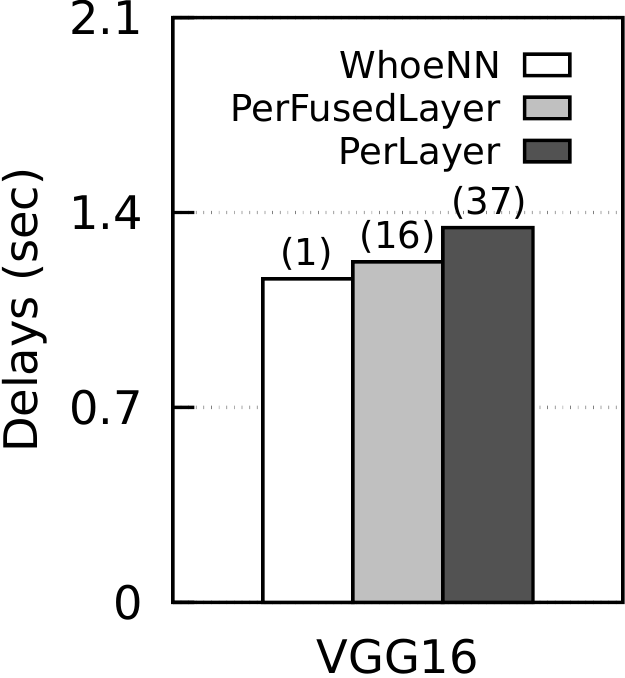}
	}
	~
	\vspace{-3pt}
	\caption{NN inference delays (including startup) with various granularities. The count of recordings is annotated.}
	\label{fig:granularity}
\end{figure}

\paragraph{Impact of recording granularity}
% For NN inference, recording every several layers is a useful tradeoff between composability and efficiency.
We tested three granularities: one monolithic recording per NN (high efficiency); one recording per NN layer (high composability); 
per fused layer with layer fusion done by ACL~\cite{acl} (a middle ground). 
%one recording per  convolutional layer and the subsequent layers of pooling and \Note{XXX}.
%For the last, we manually decide recording boundaries so that each recording is aligned with fused layers by ACL, e.g. a convolutional layer and the subsequent layers of activation, normalization, and pooling~\cite{acl}.
Figure~\ref{fig:granularity} shows that recordings of fused layers incur only 15\% longer delays on average than a monolithic recording.  
% they incur 11\% shorter delays on average than per-layer recordings. 
% replaying per-job recordings incurs delays longer by 5\% -- 17\%. 
The additional delays come from replayer startup (see Figure~\ref{fig:perf-ml-startup}). % for each recording. 
% resetting GPUs, setting page tables, and input/output injection/extraction. 
% They are modest because the startup is cheap (see Figure~\ref{fig:perf-ml-startup}). 
We conclude that for NN inference, recording every fused layer is a useful tradeoff between composability and efficiency.
% modest, because our replayer takes 
% The primary cause is that the replayer pays a startup cost for each job as opposed to an entire NN: resetting GPUs, setting up page tables, and inject the input.
% Such memory copy overhead is pronounced on compute-intense NNs, e.g. Res12 and VGG16, because the overhead is amortized over longer execution of individual jobs. 

%The primary slowdown cause is the replayer has to inject/extract the input/output for each job, which moves more data between between the memories of CPU and GPU. 
%For instance, XXX MBs of extra data is done for MobileNet. 
%\Note{Explain why sqz sees higher overhead}
%Such memory copy overhead is pronounced on compute-intense NNs, e.g. Res12 and VGG16, because the overhead is amortized over longer execution of individual jobs. 

%\jin{possible overhead: remapping intermediate data - requires pgt update / GPU init (clean state, pgt population) / no benefit of high-level graph optimization -- may require more computation, we may not have mem sync overhead}

%\Note{do we have evidence that individual kernel execution is faster?}￩

% However, \rnr{} entire network is beneficial regarding performance by i) runtime graph optimization (e.g. ), ii) removing the copy overhead between job transition and from additional GPU contexts, and iii) no CPU-GPU switching during entire computation.

\paragraph{Preemption delay for interactiveness} (\S\ref{sec:replay:handoff})
We measure the delay perceived by an interactive app when it requests to preempt GPU from the replayer. 
%Such a delay is key to app interactiveness.
On both tested GPUs, the delay is below 1 ms, which translates to minor performance degradation, e.g. loss of 1 FPS for a 60 FPS app.
%Table XXX shows our measurement on XXX GPUs, the preemption delay is as low as 1 ms, which result in minor performance degradation to the interaction apps, e.g. loss of 1 to 2 frames (assuming 60 FPS). 
The reason is preemption simplicity: 
% the replayer purges GPU cache without saving any GPU memory state; 
% the replayer drops GPU state without preserving any cache; 
a preemption primarily flushes GPU cache and GPU TLB followed by a GPU soft reset. 
% Thanks to modern GPU hardware

% !TeX root = main.tex

% 07/30/2021 heejin
% For ASPLOS template, subfigure doesn't work
\begin{comment}
\begin{figure}
	\centering
%	\begin{subfigure}[b]{0.35\textwidth}
%		\includegraphics[width=\textwidth]{./figs/plot/granularity/legend.png}
%	\end{subfigure}	
%	\vspace{1pt}
	\begin{subfigure}[b]{0.135\textwidth}
		\includegraphics[width=\textwidth]{./figs/plot/granularity/mnist}
	\end{subfigure}
	~
	\hspace{3pt}
		\begin{subfigure}[b]{0.145\textwidth}
		\includegraphics[width=\textwidth]{./figs/plot/granularity/alexnet}
	\end{subfigure}
	~
	\hspace{3pt}
	\begin{subfigure}[b]{0.145\textwidth}
		\includegraphics[width=\textwidth]{./figs/plot/granularity/vgg16}
	\end{subfigure}
	\vspace{-10pt}		% use as needed
	\caption{NN inference delays (including startup) with various granularities. The count of recordings is annotated.}
	\label{fig:granularity}
	\vspace{-10pt}		% use as needed
\end{figure}
\end{comment}

%\input{fig-checkpoint}
\paragraph{Checkpoint \& restore} (\S\ref{sec:replay:handoff})
%We tested checkpointing of GPU state which allows the replayer to resume from GPU preemption;
Our results show that GPU state checkpointing is generally inferior to re-executing the whole replay. % without any checkpointing; 
%Checkpoints save time over re-executing only when they are sparse (e.g. every a few seconds) and the preemption is very frequent. 
For instance, MobileNet making one checkpoint every 16 GPU jobs (50--60 jobs in total) slows down the whole NN execution by 8x. 
The primary cause is memory dump. 
MobileNet takes 140 ms to dump all GPU memory (51 MBs) while re-executing the NN takes only 45 ms.

\section{Related Work}
\label{sec:related}

% event-driven record and replay
% generate IO event based on the log (trace) and emulate like the IO comes from device
% inject interrupt after all the exepcted register access is emulated.

%\jin{prior work: trying to create a smaller/simpler GPU stack out of the standard one; (unverified parts can go wrong; eng difficulty - porting/partition/transplanting is hard; and there's a black box)}
%\jin{keep only GPU stack related: how much they squeeze the software stack (TCB)}

\paragraph{Record and replay} was primarily used for diagnosis and debugging~\cite{sanity, replayConfusion, rtag}.
It has been applied to mobile UI apps~\cite{reran,mobiPlay}, web apps~\cite{mahimahi}, virtual machines~\cite{reVirt}, networks~\cite{ofrewind}, and whole systems~\cite{R2}.
None of prior work has applied the idea to the CPU/GPU interactions.
Related to \sys{},
Replaying syscalls and framework calls have been popular in reverse engineering GPU runtimes~\cite{rosenzweig-m1, grate, panfrost, patrace} and reducing GPU scheduling overhead~\cite{nimble}, respectively.
Unlike them, \sys{} records at the CPU/GPU boundary and therefore achieves the goal of a lean, trustworthy replayer.

\paragraph{Refactoring GPU stacks}
To leverage TEE, 
recent works isolate part of or the whole GPU stack for security.
Sugar~\cite{sugar} subsumes a full GPU stack to an app's address space.
Graviton~\cite{graviton} pushes the function of isolation and resource management from OS to a GPU's command processor. 
Telekine~\cite{telekine} spans a GPU stack between local and cloud machines at the API boundary.
%(e.g. redirecting API calls from client to server)
%Telekine spans a GPU stack between local and cloud machines at the API boundary (e.g. XXXX)~\cite{telekine}.
HIX~\cite{HIX} ports the entire GPU stack to a secure enclave and restricts the IO interconnect.
%HIX~\cite{HIX} ports the entire GPU stack to a secure enclave for protection and also restricts the IO interconnect.
%HIX moves GPU drivers to a secure enclave for protection and also modifies the IO interconnect~\cite{HIX}. 
HETEE~\cite{hetee} instantiates dedicated hardware controller and fabric to isolate the use of GPU.
%HETEE~\cite{hetee} pools multiple GPUs in a rack and instantiates dedicated hardware controller and fabric to isolate the use of these GPUs.
While efficacy has been shown, a key drawback is the high engineering effort (e.g. deep modifications of GPU software/hardware), limited to a special hardware component (e.g. software-defined PCIe fabric) and/or likely loss of compatibility with stock GPU stacks.
%; porting XXX to a PCIe fabric XXX, ...) and/or the likely loss of compatibility with the stock GPU stacks. 
Contrasting to all the above approaches of \textit{spatial} refactoring, \sys{} can be viewed as \textit{temporal} refactoring of a GPU stack -- between the development time and the run time. 
% \sys{} thus gains engineering ease and retains compatibility with the current GPU ecosystems. 
% \sys{} shares their goal of GPU stack security
%but \textit{flattens} the GPU stack on the target device as opposed to partitioning it. 
%Visor ... 

%Note: VM migration can be done by
% - suspend/resume: dump all the memory and CUDA/GPU states when there is no in-flight kernels or,
% - record and replay: record a sequence of APIs and regenerate states by replaying
%---------------------------------------------------------------------------
% "vCUDA: GPU-Accelerated High-Performance Computing in Virtual Machines"
%-- API remoting
%-- interposes between runtime virt GPU (vCUDA stub)
%-- VM migration (by suspend/resume). wait until the GPU finished all the jobs while queueing newly added commands, which will be executed after migration. Insight: idle CUDA/GPU (no running kernel) have a sipmle soft/hardware states.
%---------------------------------------------------------------------------
% "LoGV: Low-overhead GPGPU Virtualization"
%-- para-virtualization, hypervisor running atop host device driver.
%-- VMs have guest driver which redirect requests to the hypervisor.
%-- Hypervisor does necessary check (e.g. isolation, request forwarding to host device driver such as GPU memory allocation requests)
%-- VM migration (by suspend/resume). Same mechanism with one described by vCUDA

\paragraph{GPU virtualization} often interposes between GPU stack layers in order to intercept and forward interactions, e.g. to a hypervisor~\cite{gpuvm} or to a remote server~\cite{rcuda}.
The interposed interfaces include GPU APIs~\cite{rcuda,AvA} and GPU MMIO~\cite{svga2,gpuvm}.
Notably, AvA~\cite{AvA} records and replays API calls during GPU VM migration.
\sys{} shares the principle of interposition and gives it a new use -- for recording computations ahead of time and later replaying it on a different machine.

\paragraph{Optimizing ML on GPU}
% Much studies are done on GPU ine
Much work has optimized mobile ML, e.g. by exploiting CPU/GPU heterogeneity~\cite{ulayer}. 
Notably, recent studies found CPU's software inefficiency leaving GPU under-utilized, e.g. suboptimal CLFlush~\cite{KLO} or expensive data transformation~\cite{smaug}. 
%Notably, recent studies found GPU hardware is often under-utilized, e.g. due to suboptimal CLFlush~\cite{KLO} or expensive data transformation~\cite{smaug}. 
While prior solutions fix the causes of inefficiency in the GPU stack~\cite{KLO}, 
%\sys{} offers \textit{blind} fixes without requiring to pinpoint the causes:
\sys{} offers \textit{blind} fixes without knowing the causes:
 replaying the CPU outcome (e.g. shader code) and removing  GPU idle intervals.

\paragraph{Secure ML}
%Yubara, occulumency, darkne TZ cannot support GPU computation
%Slalom only support GPU computation
%partitioning - yubara, occulumency, darkne TZ, slalom
Much work has transformed ML workloads rather than the GPU stack;
outsourcing security-sensitive compute to TEE, they preserve data/model privacy or ensure compute integrity~\cite{ternary, darkneTZ,occlumency}. 
%For instance, by outsourcing the security-sensitive compute to TEE, they preserve data/model privacy or ensure compute integrity~\cite{ternary, darkneTZ,occlumency}. 
They often support CPU-only compute and their workload transformation is orthogonal to \sys{}.
%They are often limited on CPU-only computation and their workload transformation is orthogonal to \sys{}.
While Slalom~\cite{slalom} proposed secure GPU offloading, it requires GPU stack in TEE and is limited to linear operations.

\section{Concluding Remarks}

%\paragraph{Discussion}

% ---- useful, save for later ---- 
\begin{comment}
\paragraph{Recording on target devices} 
When an app is installed to the target, 
\sys{} records its GPU executions for future replay. 
Doing so reduces developers' efforts in some use cases, e.g. replaying within TEE, since the developers do not have to know the target GPUs. 
It does not ease deployment because the target still needs a full stack. 
It raises a challenge of ensuring trustworthiness of the GPU stack for recording, which can be addressed by isolating the stack in a virtual machine and measuring its code for integrity. 

%Assuming the target device is trustworthy (e.g. with guarantees of system integrity and offline (no remote access/attack possible), recording on target device is feasible. However, some expert knowledge is recommended (e.g. for the formal verification).

%\textit{Recording from target device}.
%Assuming the target device is trustworthy (e.g. with guarantees of system integrity and offline (no remote access/attack possible), recording on target device is feasible. However, some expert knowledge is recommended (e.g. for the formal verification).
\end{comment}

\paragraph{Broader applicability}
(1) The idea of \sys{} applies to discrete GPUs. 
Our GPU hardware assumptions (\S\ref{sec:abs_model}) see counterparts on discrete GPUs albeit in different forms, e.g. registers and memory mapped via PCIe. 
In particular, \sys{} can leverage NVIDIA MIG~\cite{nvidia-mig} to enable app multiplexing: the replayer can own an MIG instance while other apps use other instances; 
% Replayer enjoys determinism within its instance. 
they are multiplexed on a physical GPU transparently by MIG. 
However, discrete GPUs raise new challenges including more complex CPU/GPU interactions, higher GPU dynamism, and recording cost due to larger memory dumps. 
%GPURip is not designed for graphics. Unlike ML, graphics tends to (1) produce dynamic GPU kernels that can hardly be enumerated ahead of time; (2) produce smaller kernels for which fine-grained multiplexing is needed for efficiency. Details in Sec2.1. 
(2) While this paper focuses on ML workloads, 
\sys{} can extend to more GPU computation including numeric analysis and physics simulation. 
(3) \sys{}'s principle is applicable to other TEEs.
A replayer in an SGX enclave is possible, but would need additional support such as MMIO remoting or SGX's extension for MMIO~\cite{HIX} because by default enclaves cannot directly access GPU registers. 

%Although not demonstrated yet in this work, we believe the \sys{} can be deployed on discrete GPUs as share the common denominator (\S~\ref{sec:abs_model}) with the integrated one (e.g. register I/O as GPU interface);
%the only exceptional feature is in their own device memory.
%However, as discrete GPUs also read/write data based on their virtual memory, it would not be different what replayer should reconstruct page table for integrated GPUs (e.g. diff is possibly sequence of replay action? (register I/O) for mem alloc and mapping and page table reconstruction).
%\Note{no support of demanding page}.

\paragraph{Recommendation to GPU vendors}
We build \sys{} without vendor support, respecting the GPU runtime blackbox (\S\ref{sec:bkgnd}) and only reasoning/modifying at the driver level. 
It would be more attractive if vendors can implement \sys{} and maintain as part of their GPU stacks. 
On one hand, the vendors can make \sys{} more robust with first-party knowledge (e.g. GPU state machines for detecting state divergence) and lightweight interface augmentation (e.g. the runtime directly discloses a job's input/output addresses). 
On the other hand, the modifications to GPU stacks are very minor and the GPU runtime internals still remain proprietary. 

%\paragraph{Summary}
%\sys{} pre-records GPU executions for replay on new input data 
%without a GPU stack.
%\sys{} identifies key GPU/CPU interactions and memory states, 
%works around proprietary GPU internals,
%and prevents replay divergence.
%The resultant replayer is tiny, portable, and quick to launch.

\section*{Acknowledgments}
The authors were supported in part by NSF awards \#1846102, \#1919197, and \#2106893.
We thank our shepherd, Dr. Mark Silberstein, and the anonymous reviewers for their insightful suggestions.

% copied from vldb submision template

%\bibliographystyle{abbrv}
\balance
\bibliographystyle{ACM-Reference-Format}

%https://tex.stackexchange.com/questions/1522/pdfendlink-ended-up-in-different-nesting-level-than-pdfstartlink
\interlinepenalty=10000	% jin: It prevents breaking a citation across two pages.

%\bibliography{vldb_sample}  % vldb_sample.bib is the name of the Bibliography in this case
\bibliography{bib/abr-short,bib/xzl,bib/hongyu,bib/misc,bib/book,bib/security,bib/iot,bib/datacentric,bib/hp,bib/numa,bib/ml-edge,bib/secureGPU,bib/TrustZone,bib/tzgpu}


%\small	
%\bibliographystyle{plain}
%\bibliographystyle{../bib/abbrv-recg}
%\bibliographystyle{bib/abbrv-minimal}
%\bibliography{bib/abr-short,bib/xzl,bib/hongyu,bib/misc,bib/book,bib/security,bib/iot,bib/datacentric,bib/hp}

%\balancecolumns
% That's all folks!
\end{document}